\documentclass[twocolumn,aps,prd,groupedaddress,amsmath,floatfix,nofootinbib]{revtex4-2}
\usepackage[colorlinks,linkcolor=blue,hyperfootnotes=true]{hyperref}
\usepackage{amsmath}
\usepackage{graphicx}% Include figure files
\usepackage{dcolumn}% Align table columns on decimal point
\usepackage{bm}% bold math
\usepackage{footnote}
\usepackage{extarrows}
\usepackage{longtable}
\usepackage{threeparttable}
\usepackage{tabularx}
\usepackage{adjustbox}
\usepackage{amssymb}

\allowdisplaybreaks[2]

\begin{document}

%\preprint{}

\title{A General Discussion on Photon Spheres in Different Categories of Spacetimes}

\author{Chen-Kai Qiao}
\email{Email: chenkaiqiao@cqut.edu.cn}
\affiliation{College of Science, Chongqing University of Technology, Banan, Chongqing, 400054, China}

\author{Ping Su}
\email{Email: suping@cqut.edu.cn}
\affiliation{College of Science, Chongqing University of Technology, Banan, Chongqing, 400054, China}

\author{Yang Huang}
\email{Email: yanghuang@mail.bnu.edu.cn}
\affiliation{School of Physics and Electronic Science, Hunan University of Science and Technology, Xiangtan, Hunan, 411201, China}

\date{\today}

\begin{abstract}
Photon spheres have attracted considerable interest in the studies of black holes and other astrophysical objects. For different categories of spacetimes (or gravitational sources), the existence of photon spheres and their distributions are dramatically influenced by the geometric and topological properties of spacetimes and characteristics of the corresponding gravitational fields. In this work, we carry out a geometric analysis on photon spheres for different categories of spacetime (including black hole spacetime, ultra-compact object's spacetime, regular spacetime, and naked singularity spacetime). Some universal properties and conclusions are obtained for these spacetimes. We mostly focus on the existence of photon spheres, the total number of photon spheres $n = n_{\text{stable}} + n_{\text{unstable}}$, the subtraction of stable photon sphere and unstable photon sphere $w = n_{\text{stable}} - n_{\text{unstable}}$ in different categories of spacetimes. These conclusions are derived solely from geometric properties of optical geometry of spacetimes, irrelevant to the specific spacetime metric forms. Besides, our results successfully recover some important theorems on photon spheres proposed in recent years.
\end{abstract}


\maketitle

\section{Introduction \label{section1}}

The photon spheres / circular photon orbits have attracted considerable interest in physics and astronomy since the capture of black hole images by Event Horizon Telescope (EHT) \cite{EHT2019a,EHT2019b,ETH2022} and the observation of gravitational waves by LIGO and Virgo \cite{LIGO2016,LIGO2016b}. Firstly, photon spheres have extremely large influences on a number of astrophysical observations, such as gravitational lensing \cite{Virbhadra2000,Bozza2002,Iyer2006,Tsukamoto2016}, black hole shadows \cite{Grover2017,Tsukamoto2018,Gralla2019,Perlick2022,Vagnozzi2023,Chen2023}, astrophysical accretion disks \cite{Jaroszynski1997,Gralla2019,ZengXX2020},  quasi-normal modes \cite{Cardoso2008}, nonlinear stability of spacetime \cite{Cardoso2014,Keir2016,GaoSJ2022,Cunha2023}, and chaotic behavior of gravitational systems \cite{Cardoso2008,Deich2023}. Furthermore, photon spheres also have the ability to reveal the underlying physical properties and characteristics of gravitational sources \cite{Hod2011,Hod2013,Johannsen2013,Mishra2019,Wielgus2021,Raffaelli2021,Bargueno2023,Vertogradov2024}, providing valuable insights into gravity theories. A number of recent studies suggested that the number and distribution of photon spheres in different categories of spacetimes (such as black hole spacetimes, ultra-compact object's spacetimes, regular spacetimes, and naked singularity spacetimes) may exhibit entirely different features \cite{Shaikh2018,Joshi2020,Berry2020,Cunha2017,Cunha2020,WeiSW2020,Isomura2023,Tsukamoto2024,Murk2024}.

The existence of photon spheres around the gravitational sources and their distribution characteristics are of profound significance. In the classical general relativity (GR) solutions, the Schwarzschild, Reissner-Nordstrom(RN), Kerr black holes all admit a single unstable photon sphere / light ring \footnote{For Kerr black holes, one may be confused from this description, because of the co-rotating and counter-rotating of photon orbits. However, If the sign of angular momentum for the photon orbit is fixed, there is a unique unstable light ring in the equatorial plane.}. On the other hand, hairy black hole solutions coupled with scalar matter fields and electromagnetic fields could admit multiple photon spheres \cite{LiuHS2019,GanQY2021,GanQY2021b,GuoGZ2023}. For black hole systems, in agreement with the capture of shadow images in astrophysical observations, it is widely postulated that the existence of unstable photon sphere must be a general feature of black hole spacetimes. Recently, a number of scholars proved the existence of unstable photon spheres in the vicinity of black holes, based on several different methods \cite{Cunha2020,WeiSW2020,Cederbaum2015,Cederbaum2016,JiaJJ2018a,JiaJJ2018b,Gibbons2016,Cunha2020,GaoSJ2021,Ghosh2021}. However, in other categories of spacetime (such as ultra-compact object's spacetime, regular spacetime, and naked singularity spacetime), the existence of photon spheres may become subtle. For instance, the studies on naked singularity spacetime shown that the unstable photon spheres can disappear in such spacetimes \cite{Shaikh2018,Joshi2020}. The presence of photon spheres in regular spacetime is more complex than in conventional black hole spacetime, which could exhibit interesting behaviors \cite{Isomura2023,Murk2024,Berry2020}. Furthermore, it is also suggested that the presence and absence of event horizon could have strong influences on the number of photon spheres in gravitational spacetimes \cite{Cunha2017,Cunha2020}. Under such circumstances, a general and comprehensive discussion on photon spheres in different categories of spacetime is necessary and urgent, which could provide us a pathway to distinguish different kinds of spacetimes theoretically (or observationally) and deepen our understanding of these spacetimes.

The explorations of photon spheres in gravitational fields can be conducted using a variety of methods. Traditionally, the positions of photon spheres / light rings near gravitational sources are often obtained through the effective potential of photons. However, the prosperity of modern geometry and topology provides us with new approaches and techniques for the studies of gravitational theories and photon spheres. Recently, fancy approaches have emerged to explore the photon spheres / light rings in a number of gravitational systems \cite{Cunha2017,Cunha2018,Cunha2020,WeiSW2020,Virbhadra2001,Koga2019,Kobialko2022,SongY2023}. Particularly, P. V. P. Cunha \emph{et al.} proposed a topological approach to photon spheres, assigning a topological invariant to each photon sphere /light ring using auxiliary vector fields \cite{Cunha2017}. Within this approach, they proved that the subtraction of stable and unstable photon spheres satisfies $n_{\text{stable}} - n_{\text{unstable}} = -1$ in black hole spacetimes \cite{Cunha2020}, which provides a demonstration on the existence of unstable photon spheres in black hole spacetimes. S. W. Wei reformulated this topological invariant as the topological charge in a recent work \cite{WeiSW2020}, using the $\phi$-mapping topological current proposed by Y. S. Duan \emph{et al} \cite{DuanYS}. Subsequently, this topological approach has been generalized into a wider range of gravitational systems and various circular orbits, including the null circular geodesics for photons and timelike circular orbits for massive particles \cite{WeiSW2023,WeiSW2023b,WeiSW2023c,YinJ2023,Lima2022,Tavlayan2023,Sadeghi2024,Cunha2024,Xavier2024,Afshar2024,Afshar2024b}. Notably, S. W. Wei \emph{et al.} suggested that the topological charge can further be used to study black hole topology and black hole thermodynamic transitions \cite{WeiSW2022,WeiSW2022b,WeiSW2024,WuD2024}. 

Inspired by the aforementioned studies, we proposed a geometric approach to study photon spheres for spherically symmetric spacetimes \cite{QiaoCK2022a,QiaoCK2022b}. In this geometric approach, the photon spheres are explored through the optical geometry of 4-dimensional Lorentz spacetime. The construction of optical geometry can be viewed as a generalization of the renewed Fermat's principle into curved spacetimes \cite{Abramowicz1988,Gibbons2008,Gibbons2009}. In 2-dimensional optical geometry, two intrinsic curvatures (the Gaussian curvature and geodesic curvature) completely determine the stable and unstable photon spheres. The proper location of photon spheres is uniquely constrained by geodesic curvature (via $\kappa_{g}(r_{ph})=0$), and the sign of Gaussian curvature determines the stability of photon spheres. Particularly, the negative Gaussian curvature suggests that photon sphere is unstable, while the positive Gaussian curvature indicates photon sphere is stable. It is also proved that our geometric approach obtains completely equivalent results with the conventional approach (using effective potentials) and the topological approach on photon spheres \cite{QiaoCK2022a,QiaoCK2022b,Cunha2022}. The geometric viewpoints behind our geometry approach have stimulated several studies in related fields \cite{Cunha2022,Bermudez-Cardenas2024,Gallo2024}, such as the extension into timelike circular geodesics and particle surfaces (in which the curvatures in Jacobi geometry of spacetimes are utilized) \cite{Cunha2022}. Furthermore, based on a geometric analysis in optical geometry, we successfully gave a new proof on the relation $n_{\text{stable}} - n_{\text{unstable}} = -1$ in black hole spacetimes \cite{QiaoCK2024}. A natural question arises: whether the characteristics of photon sphere distributions in other categories of spacetimes (e.g., ultra-compact object's spacetimes, regular spacetimes, and naked singularity spacetimes) can be analyzed and derived in a similar manner.

In the current work, we carry out a comprehensive analysis on photon spheres in several categories of spacetimes (including black hole spacetime, ultra-compact object's spacetime, regular spacetime, naked singularity spacetime), using our geometric approach in references \cite{QiaoCK2022a,QiaoCK2022b,QiaoCK2024}. Assuming the most common asymptotic behaviors of spacetime (the asymptotically flat, asymptotically de-Sitter and asymptotically anti de-Sitter), the existence of photon spheres, total number of photon spheres $n = n_{\text{stable}} + n_{\text{unstable}}$, the distribution features of stable and unstable photon spheres, and the subtraction of stable photon spheres and unstable photon spheres $w = n_{\text{stable}} - n_{\text{unstable}}$ (which can be viewed as a topological charge / topological invariant in different kinds of spacetime), are studied and discussed in details for these categories of spacetimes. This analysis would enable us to extract some universal and common features of photon spheres in different categories of spacetimes. Furthermore, it should be mentioned that our geometric analysis in the presented work is suitable to general (static) spherically symmetric spacetimes, with spacetime metric to be $ds^{2}=g_{tt}dt^{2}+g_{rr}dr^{2}+g_{\theta\theta}d\theta^{2}+g_{\phi\phi}d\phi^{2}$, irrespective of any specific forms of spacetime metric components.

The rest of the present work is organized as follows. Section \ref{section2} concisely reviews our geometric approach to photon spheres. Section \ref{section3} presents the discussions on the existence of photon spheres in different categories of spacetimes. Section \ref{section4} studies the distribution properties of stable and unstable photon spheres, especially the subtraction of stable and unstable photon spheres $w = n_{\text{stable}} - n_{\text{unstable}}$. Section \ref{section5} discusses the possible observational methods to give a constraint on photon spheres. The conclusions and perspectives are summarized in section \ref{section6}. 

\section{Geometric Approach to Photon Spheres \label{section2}}

In this section, we give a concise review about the geometric approach to photon spheres, which was proposed in recent works \cite{QiaoCK2022a,QiaoCK2022b,QiaoCK2024}. In this approach, the low-dimensional analog of Lorentz spacetime --- the optical geometry --- is employed to analyze photon spheres. In spherically symmetric gravitational systems, two intrinsic curvatures (geodesic curvature and Gaussian curvature) in optical geometry are capable of determining the locations of photon spheres and their stability. 

\begin{table*}
	\caption{Features of the conventional effective potential approach, the topological approach, and our geometric approach.}
	\label{table1}
	\vspace{1mm}
	\begin{ruledtabular}
		\begin{tabular}{lccccc}
			\\ [-7pt]
			Approach & Our Geometric Approach & Topological Approach & Conventional Approach &
			\\ [3pt]
			\hline
			\\ [-7pt]
			Geometry & Optical Geometry       & Spacetime Geometry   & Spacetime Geometry &
			\\ [2pt]
			\hline
			\\ [-7pt]
			Basic quantities & Gaussian Curvature $\mathcal{K}(r)$ & Auxiliary Vector Field $V$ & Effective Potential $V_{\text{eff}}(r)$ &
			\\ 
			& Geodesic Curvature $\kappa_{g}(r)$ & Topological Charge $w$ &       
			\\ [3pt]
			\hline
			\\ [-7pt]
			Photon Sphere  & $\kappa_{g}(r)=0$ & $V=0$ & $\frac{dV_{\text{eff}}(r)}{dr}=0$
			\\
			Unstable Photon Sphere & $\kappa_{g}(r)=0$ and $\mathcal{K}(r)<0$ & $V=0$ and $w=-1$ & $\frac{dV_{\text{eff}}(r)}{dr}=0$ and $\frac{d^{2}V_{\text{eff}}(r)}{dr^{2}}<0$ &
			\\
			Stable Photon Sphere & $\kappa_{g}(r)=0$ and $\mathcal{K}(r)>0$ & $V=0$ and $w=+1$ & $\frac{dV_{\text{eff}}(r)}{dr}=0$ and $\frac{d^{2}V_{\text{eff}}(r)}{dr^{2}}>0$ &
			\\ [3pt]
		\end{tabular}
	\end{ruledtabular}
\end{table*}

Mathematically, the optical geometry can be constructed in several equivalent ways \cite{Abramowicz1988,Gibbons2008,Gibbons2009}. A simple approach to realize the optical geometry is from a continuous map of the spacetime geometry using the null constraint $d\tau^{2}=-ds^{2}=0$ 
\begin{equation}
	\underbrace{ds^{2} = g_{\mu\nu}dx^{\mu}dx^{\nu}}_{\text{Spacetime Geometry}}
	\ \ \overset{d\tau^{2}=-ds^{2}=0}{\Longrightarrow} \ \ 
	\underbrace{dt^{2} = g^{\text{OP}}_{ij}dx^{i}dx^{j}}_{\text{Optical Geometry}}
	\label{optical geometry}
\end{equation}
For any spherically symmetric spacetimes, the photon orbits can always be constricted in the equatorial plane without loss of generality. In such circumstances, a 2-dimensional optical geometry can be constructed.
\begin{equation}
	\underbrace{dt^{2} = g^{\text{OP}}_{ij}dx^{i}dx^{j}}_{\text{Optical Geometry}}
	\ \ \overset{\theta=\pi/2}{\Longrightarrow} \ \ 
	\underbrace{dt^{2}=\tilde{g}^{\text{OP-2d}}_{ij}dx^{i}dx^{j}}_{\text{Optical Geometry (Two Dimensional)}}
	\label{optical geometry2}
\end{equation}
Particularly, in spherically symmetric spacetimes, the corresponding optical geometry $dt^{2}=\tilde{g}^{\text{OP-2d}}_{ij}dx^{i}dx^{j}$ is always described by 2-dimensional Riemannian geometry \cite{Gibbons2008,Gibbons2009,Ishihara2016a,Ishihara2016b}. The optical geometry proves to be extremely powerful in the study of gravitational deflection and gravitational lensing \cite{Gibbons2008,Gibbons2009,Werner2012,Ishihara2016a,Ishihara2016b,Asida2017,Jusufi2017,Crisnejo2018,Takizawa2020,Ono2019,LiZH2019a,LiZH2019b,LiZH2020,HuangY2022,HuangY2023,Takizawa2023,ZhangZ2021,ZhangZ2024}.

In our geometric approach, the analysis of photon spheres and their stability is carried out using intrinsic curvatures in 2-dimensional optical geometry. The most important intrinsic curvatures in 2-dimensional optical geometry are Gaussian curvature and geodesic curvature. Based on these intrinsic curvatures, the locations of photon spheres and their stability can be completely determined. Firstly, photon spheres could maintain their geodesic nature when transformed in the 2-dimensional optical geometry, so the geodesic curvature for photon sphere vanishes naturally \cite{QiaoCK2022a,QiaoCK2022b}
\begin{equation} 
	\kappa_{g}(r=r_{ph})  =  0
\end{equation}
Further, it has been proven that the geodesic curvature condition in our geometric approach is totally equivalent to the effective potential condition in conventional approach \cite{QiaoCK2022a,QiaoCK2022b}.
\begin{equation} 
	\kappa_{g}(r=r_{\text{ph}})  =  0  
	\ \ \Leftrightarrow \ \
	\frac{dV_{\text{eff}}(r)}{dr} \bigg|_{r=r_{\text{ph}}} = 0 
\end{equation}
Secondly, the stability of photon spheres can be constrained by conjugate points in optical geometry. For stable and unstable photon spheres, the behaviors of photon orbits that undergo a perturbation are completely different. When photon orbits get perturbed from unstable photon spheres, they would move far away and never return to unstable photon spheres. Conversely, when photon orbits are perturbed from stable photon spheres, they could also form other bound orbits near stable photon spheres. Mathematically, these distinct features of stable and unstable photon spheres are reflected by conjugate points in the manifold. There are conjugate points in the stable photon sphere, while no conjugate points exist in the unstable photon sphere \cite{QiaoCK2022a,QiaoCK2022b,QiaoCK2024}. The presence and absence of conjugate points provide us with a novel scheme to distinguish the stable and unstable photon spheres. 
The Cartan-Hadamard theorem in differential geometry and topology strongly constrains the Gaussian curvature and the existence of conjugate points \cite{Berger1988}. Applying the Cartan-Hadamard theorem in optical geometry, the following Gaussian curvature condition on the stability of photon spheres is derived \cite{QiaoCK2022a,QiaoCK2022b}.
\begin{eqnarray}
	\mathcal{K}(r) < 0 & \Rightarrow & \text{The photon sphere $r=r_{ph}$ is unstable} \nonumber 
	\\
	\mathcal{K}(r) > 0 & \Rightarrow & \text{The photon sphere $r=r_{ph}$ is stable} \nonumber
\end{eqnarray}
This Gaussian curvature condition, derived purely from the curvatures and topology of optical geometry, is independent of the specific metric forms of the particular gravitational systems. Consequently, the conclusion can be universally applied to any spherically spherical system. Furthermore, this Gaussian curvature condition for stable and unstable photon spheres is equivalent to the effective potential condition in conventional approach  \cite{QiaoCK2022a,QiaoCK2022b}
\begin{subequations} 
\begin{eqnarray}
		\frac{d^{2}V_{\text{eff}}(r)}{dr^{2}} \bigg|_{r=r_{\text{unstable}}} < 0 
		\ \ & \Leftrightarrow & \ \ 
		\mathcal{K}(r=r_{\text{unstable}}) < 0 \ \ \ \ \ \ \ \  %\nonumber
		\\
		\frac{d^{2}V_{\text{eff}}(r)}{dr^{2}} \bigg|_{r=r_{\text{stable}}} > 0 
		\ \ & \Leftrightarrow & \ \ 
		\mathcal{K}(r=r_{\text{stable}}) > 0
\end{eqnarray}
\end{subequations}
The distinguishing features of our geometric approach, conventional approach, and topological approach (together with the equivalence between these approaches) have been summarized in table \ref{table1}.

\section{The Existence of Photon Spheres \label{section3}}

In this section, we present a discussion on the existence of photon spheres in general spherically symmetric gravitational systems, including black hole spacetimes, ultra-compact objects’ spacetimes, regular spacetimes (free of spacetime singularities), and naked singularity spacetimes.  

The general (static) spherically symmetric spacetime metric can be expressed as
\begin{equation}
	ds^{2} = g_{tt}(r) dt^{2} + g_{rr}(r) dr^{2} 
	+ g_{\theta\theta}(r) d\theta^{2} 
	+ g_{\phi\phi}(r,\theta) d\phi^{2}
	\label{spacetime metric}
\end{equation}
Additionally, for any spherically symmetric spacetimes, the metric components $g_{\theta\theta}=r^{2}$ and $g_{\phi\phi}=r^{2}\sin^{2}\theta$ can always be achieved through a coordinate transformation, which makes the spacetime metric reduced to be
\begin{equation}
	ds^{2} = - f(r) dt^{2} + g(r) dr^{2} 
	+ r^{2} d\theta^{2} + r^{2} \sin^{2}\theta d\phi^{2}
	\label{spacetime metric reduced0}
\end{equation}
with $f(r)=-g_{tt}(r)$ and $g(r)=g_{rr}(r)$. The corresponding 2-dimensional optical geometry (restricted in the equatorial plane $\theta=\pi/2$) of spherically symmetric spacetime is given by 
\begin{equation}
	dt^{2} = \tilde{g}^{\text{OP-2d}}_{ij}dx^{i}dx^{j} 
	% & = & \tilde{g}^{\text{OP-2d}}_{rr}dr^{2} + \tilde{g}^{\text{OP-2d}}_{\phi\phi}d\phi^{2} \nonumber
	% \\
	= \frac{g(r)}{f(r)} \cdot dr^{2} + \frac{r^{2}}{f(r)} \cdot d\phi^{2}
\end{equation} 
The geodesic curvature of a circular curve with constant radius $r$ (such as photon spheres) in this 2-dimensional optical geometry can be calculated via
\begin{eqnarray}
	\kappa_{g}(r) 
	& = & \frac{1}{2\sqrt{\tilde{g}^{\text{OP-2d}}_{rr}}} 
	\frac{\partial \big[ \text{log}(\tilde{g}^{\text{OP-2d}}_{\phi\phi})\big]}{\partial r}
	\nonumber
	\\ 
	& = & \frac{1}{\sqrt{f(r) \cdot g(r)}} 
	\bigg[ \frac{f(r)}{r} - \frac{1}{2} \frac{df(r)}{dr} \bigg]
	\label{geodesic cuurvature expression}
\end{eqnarray}
\begin{widetext} 
and the Gaussian curvature in 2-dimensional optical geometry can be calculated through
	\begin{eqnarray}
		\mathcal{K} 
		& = & -\frac{1}{\sqrt{\tilde{g}^{\text{OP-2d}}}} 
		\bigg[
		\frac{\partial}{\partial \phi} \bigg( \frac{1}{\sqrt{\tilde{g}^{\text{OP-2d}}_{\phi\phi}}} \frac{\partial\sqrt{\tilde{g}^{\text{OP-2d}}_{rr}}}{\partial \phi}  \bigg)
		+ \frac{\partial}{\partial r} \bigg( \frac{1}{\sqrt{\tilde{g}^{\text{OP-2d}}_{rr}}} \frac{\partial\sqrt{\tilde{g}^{\text{OP-2d}}_{\phi\phi}}}{\partial r}  \bigg)
		\bigg]  \nonumber
		\\
		& = &   \frac{1}{2r} \frac{1}{g(r)} \frac{df(r)}{dr} 
		+ \frac{1}{2r} \frac{f(r)}{[g(r)]^2} \frac{dg(r)}{dr} 
		- \frac{1}{2f(r) g(r)} \bigg[ \frac{df(r)}{dr} \bigg]^2
		- \frac{1}{4 [g(r)]^2} \frac{df(r)}{dr} \frac{dg(r)}{dr} 
		+ \frac{1}{2g(r)} \frac{d^2 f(r)}{dr^2}
		\label{Gaussian cuurvature expression}
	\end{eqnarray}
\end{widetext}
with $\tilde{g}^{\text{OP-2d}}=\text{det} \big( \tilde{g}^{\text{OP-2d}}_{ij} \big)$ to be the determinant of the 2-dimensional optical geometry metric. 
In the present work, all conclusions on photon spheres are obtained from geometric analysis (using geodesic curvature in optical geometry) and the asymptotic behaviors of gravitational systems, irrespective of any specific spacetime metric functions $g_{tt}$, $g_{rr}$, $g_{\theta\theta}$ and $g_{\phi\phi}$.

From the geometric approach described in section \ref{section2}, the condition for photon spheres is the vanishing of geodesic curvature, namely $\kappa_{g}(r_{ph}) = 0$. Therefore, the existence of photon spheres suggests that equation $\kappa_{g}(r) = 0$ admits at least one solution \cite{QiaoCK2024}. To provide an analysis on the existence of photon spheres in general spherically symmetric gravitational systems, the crucial point is studying the behavior of geodesic curvature $\kappa_{g}(r)$ for a circular curve in the 2-dimensional optical geometry.

\begin{figure*}
	\includegraphics[width=0.95\textwidth]{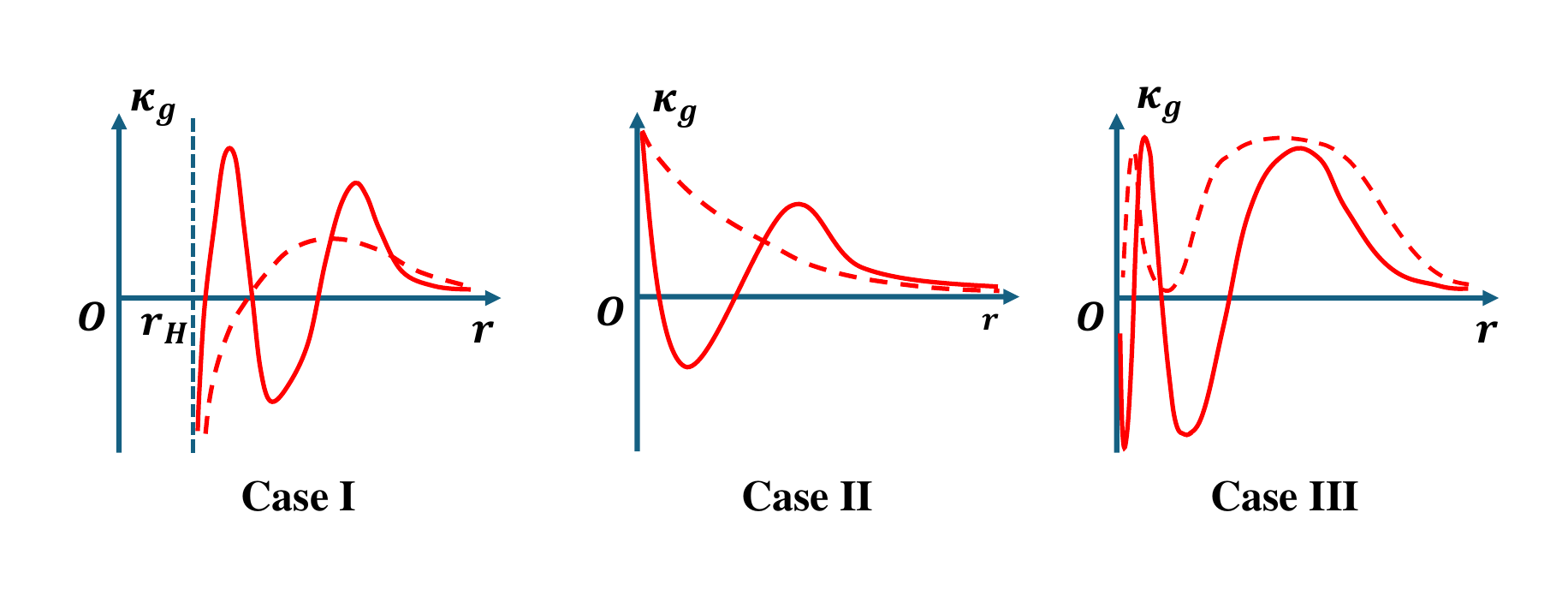}
	\caption{The variation of geodesic curvature $\kappa_{g}(r)$ with respect to radial coordinate $r$ in different categories of spacetimes. \textbf{Case I:} For black hole spacetime and regular spacetime (with the presence of horizons) satisfying $\lim_{r \to r_{H}} \kappa_{g}(r) < 0$ and $\lim_{r \to \infty} \kappa_{g}(r) = 0^{+}$, the equation $\kappa_{g}(r)=0$ must have at least one solution outside the event horizon. The dashed line draws the one-solution case, and the solid line draws the three-solution case. \textbf{Case II:} For ultra-compact object’s spacetime, naked singularity spacetime (with finite first-order metric derivative), and regular spacetime (without horizons) satisfying $\lim_{r \to 0} \kappa_{g}(r) > 0$ and $\lim_{r \to \infty} \kappa_{g}(r) = 0^{+}$, the equation $\kappa_{g}(r)=0$ must have either no solution or an even number of solutions. The dashed line illustrates the no-solution case, and the solid line draws the two-solution case. \textbf{Case III:} For naked singularity spacetime (with first-order metric derivative diverging to positive infinity at spacetime singularity $\lim_{r\to 0}\frac{df(r)}{dr} = +\infty$), the geodesic curvature satisfies $\lim_{r \to 0} \kappa_{g}(r) = \text{indefinite}$ and $\lim_{r \to \infty} \kappa_{g}(r) = 0^{+}$. It is impossible to determine whether equation $\kappa_{g}(r)=0$ has solutions or not. The solid line depicts an example where the solution of $\kappa_{g}(r)=0$ exists, while the dashed line illustrates an example where $\kappa_{g}(r)=0$ has no solutions. In this figure, the notation $\lim_{r\to\infty}\kappa_{g}(r)=0^{+}$ represents that the geodesic curvature satisfies $\kappa_{g}(r) \to 0$ and $\kappa_{g}(r)>0$ in the infinite distance limit $r \to \infty$.}
	\label{figure1}
\end{figure*}

For different categories of spacetime, the valid regions in which photon spheres could appear are slightly different. When spacetime has an event horizon, the valid region for the appearance of photon spheres is generally outside the horizon $r>r_{H}$ (the photon spheres inside the horizons, if they existed, are unable to be captured by any astrophysical observer outside the horizon). On the contrary, if the spacetime does not have event horizons, the positions of photon spheres may occur in the entire spacetime region $r>0$. To give a comprehensive discussion on the existence of photon spheres, the geodesic curvature $\kappa_{g}(r)$ in the inner and outer boundaries of the valid region (where photon spheres could exist) needed to be studied for different categories of spacetimes. Particularly, for black hole spacetime and regular spacetime (without singularities, but have horizons), the geodesic curvature in the near horizon region $r \to r_{H}$ and the infinite distance region $r \to \infty$ should be analyzed. For ultra-compact object’s spacetime, naked singularities, and regular spacetime (without singularity spacetime and horizons), the geodesic curvature in the $r \to 0$ limit and the infinite distance region $r \to \infty$ must be analyzed. In the present work, we shall encounter the following situations:
\begin{itemize}
	\item \textbf{CASE I}: For black hole spacetime and regular spacetime with the presence of horizons, the geodesic curvature in the infinite distance limit satisfies $\lim_{r \to \infty} \kappa_{g}(r) = 0^{+}$ and in the near horizon region satisfies $\lim_{r \to r_{H}} \kappa_{g}(r) < 0$, the equation $\kappa_{g}(r)=0$ must have at least one solution outside the event horizon. This situation is illustrated in the first case of figure \ref{figure1}.
	
	\item \textbf{CASE II:} For ultra-compact object’s spacetime (without horizons), naked singularity spacetime (with finite first-order metric derivative), regular spacetime (without horizons), the geodesic curvature in the infinite distance limit satisfies $\lim_{r \to \infty} \kappa_{g}(r) = 0^{+}$ and in center limit satisfies $\lim_{r \to 0} \kappa_{g}(r) > 0$, the equation $\kappa_{g}(r)=0$ must have either no solution or an even number of solutions. This situation is shown in the second case of figure \ref{figure1}.
	
	\item \textbf{CASE III}: For naked singularity spacetime (with first-order metric derivative diverging to positive infinity at spacetime singularity $\lim_{r\to 0}\frac{df(r)}{dr} = +\infty$), the geodesic curvature in the infinite distance limit satisfies $\lim_{r \to \infty} \kappa_{g}(r) = 0^{+}$, while the geodesic curvature in the center limit becomes indefinite ($\lim_{r \to 0} \kappa_{g}(r) = \text{indefinite}$). Under this circumstance, it is impossible to determine whether equation $\kappa_{g}(r)=0$ has solutions or not. This situation has been shown in the third case of figure \ref{figure1}.
\end{itemize} 
In the rest of this section, we analyze the behavior of geodesic curvature $\kappa_{g}(r)$ for each case of figure \ref{figure1} in different categories of spacetimes independently. 

\subsection{Geodesic Curvature in the Infinite Distance Limit $r \to \infty$  \label{section3a}}

This subsection discusses the behavior of geodesic curvature $\kappa_{g}(r)$ in the infinite distance region $r \to \infty$ in the 2-dimensional optical geometry. As we will see in the following procedures, the geodesic curvature in the infinite distance limit only depends on the asymptotic expansions of spacetime metric, irrelevant to the presence (or absence) of event horizon. The analysis and conclusion are completely the same for different categories of spacetimes (such as black hole spacetimes, ultra-compact objects’ spacetimes, regular spacetimes, and naked singularity spacetimes). In particular, the most common asymptotic behaviors of spacetimes (the asymptotically flat, asymptotically de Sitter, and asymptotically anti de Sitter) are chosen to carry out the analysis.

\textbf{Asymptotically Flat Spacetimes:}
The asymptotically flat spacetime has the following asymptotic metric expansions in the infinite distance limit $r \to \infty$
\begin{subequations}
	%\begin{eqnarray}
	\begin{align}
		\lim_{r \to \infty} f(r) & = 1 + \frac{a_{i}}{r^{i}} + O\bigg( \frac{1}{r^{i+1}} \bigg)  & (i \ge 1)
		\\
		\lim_{r \to \infty} g(r) & = \frac{1}{1 + \frac{b_{j}}{r^{j}} + O\big( \frac{1}{r^{j+1}} \big)} & (j \ge 1)
	\end{align}
	%\end{eqnarray}
\end{subequations}
with $a_{i}$ and $b_{j}$ to be expansion coefficients. Based on asymptotic metric expansions, the geodesic curvature $\kappa_{g}(r)$ in the infinite distance limit is calculated as
\begin{eqnarray}
	\lim_{r\to\infty}\kappa_{g}(r)
	& = &
	\lim_{r\to\infty} 
	\bigg[ 
	\frac{1}{\sqrt{f(r) \cdot g(r)}} 
	\bigg( \frac{f(r)}{r} - \frac{1}{2} \frac{df(r)}{dr} \bigg) 
	\bigg] \nonumber
	\\
	%& = & \lim_{r\to\infty}
	%\frac{\frac{1}{r}+\frac{a_{i}}{r^{i+1}}+\frac{ia_{i}}{2r^{i+1}}+O\big(\frac{1}{r^{i+2}}\big)}{\sqrt{\frac{1+\frac{a_{i}}{r^{i}}+O\big(\frac{1}{r^{i+1}}\big)}{1+\frac{b_{j}}{r^{j}}+O\big(\frac{1}{r^{j+1}}\big)}}} \nonumber
	%\\
	& = & \lim_{r\to\infty}
	\frac{\frac{1}{r}+\frac{a_{i}}{r^{i+1}}+\frac{ia_{i}}{2r^{i+1}}+O\big(\frac{1}{r^{i+2}}\big)}{ \sqrt{ 1 + \frac{\frac{a_{i}}{r^{i}}-\frac{b_{j}}{r^{j}}+O\big(\frac{1}{r^{i+1}}, \frac{1}{r^{j+1}}\big)}{1+\frac{b_{j}}{r^{j}}+O\big(\frac{1}{r^{j+1}}\big)}}} \nonumber
	\\
	& = & \lim_{r\to\infty} \frac{1}{r} = 0^{+} 
\end{eqnarray}
The notation $\lim_{r\to\infty}\kappa_{g}(r)=0^{+}$ represents that the geodesic curvature satisfies $\kappa_{g}(r) \to 0$ and $\kappa_{g}(r)>0$ in the $r \to \infty$ limit. This is exactly consistent with the three cases illustrated in figure \ref{figure1}.

\textbf{Asymptotically de Sitter and Asymptotically Anti de Sitter Spacetimes:}
In the infinite distance limit, the asymptotically de Sitter / asymptotically anti de Sitter spacetime has the asymptotic analytical metric expansions
\begin{subequations}
	%\begin{eqnarray}
	\begin{align}
		%\begin{flalign}
		\lim_{r \to \infty} f(r) & = 1 - \frac{\Lambda r^{2}}{3} + \frac{a_{i}}{r^{i}} + O\bigg( \frac{1}{r^{i+1}} \bigg) & (i \ge 1) 
		\\
		\lim_{r \to \infty} g(r) & = \frac{1}{1-\frac{\Lambda r^{2}}{3}+\frac{b_{j}}{r^{j}}+O\big(\frac{1}{r^{j+1}}\big)} & (j \ge 1)
		%\end{flalign}
	\end{align}
	%\end{eqnarray}
\end{subequations}
with $a_{i}$ and $b_{j}$ to be expansion coefficients and $\Lambda$ to be the cosmological constant. The asymptotically de Sitter / anti de Sitter spacetime admits a positive (or negative) cosmological constant $\Lambda>0$ (or $\Lambda<0$). Using the above metric expansions, we obtain the geodesic curvature in the infinite distance limit
\begin{eqnarray}
	\lim_{r\to\infty}\kappa_{g}(r)
	& = &
	\lim_{r\to\infty} 
	\bigg[ 
	\frac{1}{\sqrt{f(r) \cdot g(r)}} 
	\bigg( \frac{f(r)}{r} - \frac{1}{2} \frac{df(r)}{dr} \bigg) 
	\bigg] \nonumber
	\\
	%& = & \lim_{r\to\infty}
	%\frac{ \frac{1}{r} + \frac{a_{i}}{r^{i+1}} + \frac{ia_{i}}{2r^{i+1}} + O\big(\frac{1}{r^{i+2}}\big) }
	%{ \sqrt{\frac{1-\frac{\Lambda r^{2}}{3}+\frac{a_{i}}{r^{i}}+O\big(\frac{1}{r^{i+1}}\big)}{1-\frac{\Lambda r^{2}}{3}+\frac{b_{j}}{r^{j}}+O\big(\frac{1}{r^{j+1}}\big)} } } \nonumber
	%\\
	& = & \lim_{r\to\infty}
	\frac{ \frac{1}{r} + \frac{a_{i}}{r^{i+1}} + \frac{ia_{i}}{2r^{i+1}} + O\big(\frac{1}{r^{i+2}}\big) }
	{ \sqrt{ 1 + \frac{\frac{a_{i}}{r^{i}}-\frac{b_{j}}{r^{j}}+O\big(\frac{1}{r^{i+1}}, \frac{1}{r^{j+1}}\big)}{1-\frac{\Lambda r^{2}}{3}+\frac{b_{j}}{r^{j}}+O\big(\frac{1}{r^{j+1}}\big)} } } \nonumber
	\\
	& = & \lim_{r\to\infty} \frac{1}{r} = 0^{+} 
\end{eqnarray}
The notation $\lim_{r\to\infty}\kappa_{g}(r)=0^{+}$ indicates that the geodesic curvature satisfies $\kappa_{g}(r) \to 0$ and $\kappa_{g}(r)>0$ in the $r \to \infty$ limit, which is the same with results for asymptotically flat spacetimes. The geodesic curvature behavior in the infinite distance limit successfully agrees with the three cases illustrated in figure \ref{figure1}.

\subsection{Geodesic Curvature in the Near Horizon Limit $r \to r_{H}$ (or in the Center Limit $r \to 0$) \label{section3b}}

Unlike the behavior of geodesic curvature in the infinite distance limit (which depends uniquely on the asymptotic analytical expansions of metric components and is irrelevant to the horizons and types of spacetime), the geodesic curvature in the inner boundary of region (where photon spheres could exist) are strongly constrained by spacetime structures and properties of gravitational sources. For different categories of spacetimes (the black hole spacetime, ultra-compact object’s spacetime, regular spacetime, naked singularity spacetime), the geodesic curvature $\kappa_{g}(r)$ in the inner boundary are classified into two situations: the near horizon limit $r \to r_{H}$ and the center limit $r \to 0$. 

\textbf{Black Hole Spacetime:} 
According to the cosmic censorship conjecture proposed by R. Penrose \emph{et al.}, black hole spacetimes always admit event horizons, such that spacetime singularities are enclosed by event horizons \cite{Penrose1969,Hawking1973,Wald1998}. The event horizon is a null hypersurface in Lorentz spacetime and vanishes the metric component $g_{tt}(r) = -f(r)$, $g_{rr}(r)=g(r)$, and it plays significantly important roles in the causal structure of spacetime. The metric components $f(r)$ and $g(r)$ must change their signs when going across the horizon.  
For astronomical observational reasons, the locations of photon spheres are always assumed to be outside the black hole horizon, since any photon spheres inside event horizon will never be observed by a distant observer.

Notably, for black hole spacetime, a recent study has shown that the geodesic curvature in the near horizon limit is closely related to the surface gravity of black holes, via $\lim_{r \to r_{H}} \kappa_{g}(r) = - \kappa_{\text{surface gravity}} < 0$ \cite{Cunha2022}. Since the geodesic curvature in the infinite distance limit satisfies $\lim_{r \to \infty} \kappa_{g}(r) = 0^{+}$ (which has been discussed in subsection \ref{section3a}), then the equation $\kappa_{g}(r) = 0$ must have at least one solution (as illustrated in the first case of figure \ref{figure1}), confirming the existence of photon spheres in the vicinity of black holes.

However, in the present work, we apply an alternative demonstration on the existence of photon spheres in black hole spacetime, through a simpler analysis on metric functions $g_{tt}(r)=-f(r)$ and $g_{rr}(r)=g(r)$ and their derivatives near the event horizon.
Since the presence of an event horizon changes the sign of metric components $f(r)$ and $g(r)$, the metric component function near the horizon must obey the relations 
\begin{align}
	f(r) < 0 & \ \ \ \text{when} \ r < r_{H} \nonumber
	\\
	f(r) = 0 & \ \ \ \text{when} \ r \to r_{H} \nonumber
	\\
	f(r) = \text{finite} > 0 & \ \ \ \text{when} \ r > r_{H} \nonumber
\end{align}
which suggests in the near horizon limit
\begin{equation}
    \lim_{r \to r_{H}} \frac{df(r)}{dr} > 0 
    \ \ \  
    \lim_{r \to r_{H}} \frac{f(r)}{r} = 0 
    \nonumber
\end{equation}
With these relations, the geodesic curvature in 2-dimensional optical geometry in the near horizon limit becomes 
\begin{eqnarray}
	\lim_{r \to r_{H}} \kappa_{g}(r)
	& = & \lim_{r \to r_{H}} 
	\bigg[ 
	\frac{1}{\sqrt{f(r) \cdot g(r)}} 
	\bigg( \frac{f(r)}{r} - \frac{1}{2} \frac{df(r)}{dr} \bigg) 
	\bigg] \nonumber
	\\
	& = & \lim_{r \to r_{H}} 
	\bigg[ 
	\frac{1}{\sqrt{f(r) \cdot g(r)}} 
	\bigg( 0 - \frac{1}{2} \frac{df(r)}{dr} \bigg) 
	\bigg] \nonumber
	\\
	& < & 0
	\label{geodesic curvature black hole spacetime}
\end{eqnarray}
which is consistent with the result obtained by Cunha \emph{et al.} in reference \cite{Cunha2022}. In the above derivation process, it should be noted that the factor $\frac{1}{\sqrt{f(r) \cdot g(r)}}$ is assumed to be non-singular in the near horizon limit. Particularly, in the simple spherically symmetric black hole systems (such as the classical Schwarzschild black hole and RN black hole), the metric components satisfy $f(r) \cdot g(r) = 1$. In the most general spherically symmetric black hole systems, although the relationship $f(r) \cdot g(r) = 1$ is no longer valid, it is reasonable to assume that the product $f(r) \cdot g(r) > 0$ holds outside the event horizon. This can be viewed as a restriction from the causal structure of spacetime, such that the tangent vectors $\frac{\partial}{\partial t}$ and $\frac{\partial}{\partial r}$ associated with the conventional static coordinates point to the timelike and spacelike directions simultaneously outside the event horizon. 

In conclusion, for black hole spacetime, the geodesic curvature in optical geometry satisfies $\lim_{r \to r_{H}} \kappa_{g}(r) < 0$ in the near horizon limit and $\lim_{r \to \infty} \kappa_{g}(r) = 0^{+}$ in the infinite distance limit, as illustrated in the first case of figure \ref{figure1}. The existence of solutions for equation $\kappa_{g}(r)=0$ is guaranteed in such cases, demonstrating the presence of photon spheres near black holes. Furthermore, assuming the continuity of geodesic curvature $\kappa_{g}(r)$ 
\footnote{The geodesic curvature depends on the first-order derivatives of the metric components $f(r)$ and $g(r)$. The continuity of first-order derivatives of metric components $f(r)$, $g(r)$ is necessary to give the well-defined concepts of curvature tensors (or curvature scalars), both in 4-dimensional spacetime geometry and the corresponding 2-dimensional optical geometry. For spacetimes admit event horizons, their optical geometry is generally defined outside the event horizon $r \geq r_{H}$, in which the spacetime singularities (of black holes) are excluded. So we always assume the continuity of geodesic curvature in optical geometry, expect for the center point $r=0$ of the horizon-less naked singularities spacetime. \label{footnote-continuous}}, 
we can easily observe that the total number of photon spheres for black hole spacetime (outside the event horizon) must be an odd number, namely $n = n_{\text{stable}} + n_{\text{unstable}} = 2k+1$ (where $k \in \mathbb{N}$ is a natural number).

\textbf{Ultra-Compact Object’s Spacetime:} 
For spacetimes generated by ultra-compact objects (such as neutron stars, boson stars, axion stars, strange stars, and other massive compact objects as black hole mimickers), due to the absence of an event horizon, the photon sphere may emerge in the entire spacetime region $r \geq 0$ 
\footnote{Strictly speaking, due to the opacity of ultra-compact objects, only the photon spheres with a radius larger than the ultra-compact object's size ($r_{ph} \geq r_{\text{surface}}$) can be detected by distant observers in astrophysical observations. The photon spheres inside the ultra-compact objects are much too complicated, which is beyond the scope of this study. In the current work, we simply assume the size of ultra-compact objects to be sufficiently small. Under such circumstances, any photon sphere that emerges in the entire spacetime region $r \geq 0$ can be observed.}. 
For such spacetime, it is necessary to analyze the behavior of geodesic curvature $\kappa_{g}(r)$ in the center limit $r \to 0$. 

The ultra-compact objects do not produce event horizons, so the spacetime metric is regular in the entire spacetime 
\begin{align}
	f(r)=\text{finite}>0 & \ \ \ \text{when} \ r \geq 0 \nonumber
	\\
	g(r)=\text{finite}>0 & \ \ \ \text{when} \ r \geq 0 \nonumber
	\\
	\frac{df(r)}{dr} = \text{finite} & \ \ \ \text{when} \ r \geq 0 \nonumber
\end{align}
which suggests in the center limit $r \to 0$ 
\begin{equation}
	\lim_{r \to 0} \frac{f(r)}{r} = + \infty \nonumber
\end{equation}
In such cases, the geodesic curvature satisfies the following relation in the $r \to 0$ limit
\begin{eqnarray}
	\lim_{r \to 0} \kappa_{g}(r) 
	& = & \lim_{r \to 0} 
	      \bigg[ 
	        \frac{1}{\sqrt{f(r) \cdot g(r)}} 
	        \bigg( \frac{f(r)}{r} - \frac{1}{2} \frac{df(r)}{dr} \bigg) 
	      \bigg] \nonumber
	      \\
    & = & \lim_{r \to 0} 
          \bigg[ 
            \frac{1}{\sqrt{f(r) \cdot g(r)}}
            \bigg( +\infty - \text{finite} \bigg) \nonumber
          \\
    & = & + \infty     
	      \label{geodesic curvature Ultra-Compact Object’s Spacetime}
\end{eqnarray}
Therefore, for spacetimes generated by ultra-compact objects, the geodesic curvature satisfies $\lim_{r \to 0} \kappa_{g}(r) = + \infty $ in the center limit and $\lim_{r \to \infty} \kappa_{g}(r) = 0^{+}$ in the infinite distance limit (the latter has been explained in subsection \ref{section3a}), this is precisely the scenario depicted in the second case of figure \ref{figure1}. In such cases, the equation $\kappa_{g}(r)=0$ possesses either no solution or an even number of solutions \footnote{We have assumed the continuity of geodesic curvature, as explained in footnote \ref{footnote-continuous}}. 
Consequently, the existence of photon spheres is not necessary for ultra-compact object's spacetimes (such as spacetime generated by neutron stars, boson stars, axion stars, strange stars, and other massive compact objects as black hole mimickers). Furthermore, for ultra-compact object's spacetimes, the total number of photon spheres must be even, namely $n = n_{\text{stable}} + n_{\text{unstable}} = 2k$ (where $k \in \mathbb{N}$ is a natural number).

\textbf{Regular Spacetime (with the presence of event horizon):} 
In this part, we analyze the regular spacetimes, which are free of spacetime singularities but with the presence of event horizons. Similar to the black hole cases, we only study  photon spheres that are located outside the event horizon for astronomical observational reasons. Although a recent study shows that photon spheres may exist inside the event horizon for special regular spacetime \cite{Isomura2023}, their connections with observations are still greatly challenged. Therefore, we only focus on photon spheres outside the event horizons, the discussion of photon spheres inside the horizon is beyond the scope of our present study.

For regular spacetime that possesses a horizon, since the presence of an event horizon changes the sign of metric components $f(r)$ and $g(r)$, the metric components near and outside the horizon must obey the relation 
\begin{align}
	\lim_{r \to r_{H}} f(r) = 0 & \ \ \ \text{when} \ r \to r_{H} \nonumber
	\\
	f(r) = \text{finite} > 0 & \ \ \ \text{when} \ r > r_{H} \nonumber
	\\
	g(r) = \text{finite} > 0 & \ \ \ \text{when} \ r > r_{H} \nonumber
\end{align}
which suggests in the near horizon limit $r \to r_{H}$
\begin{equation}
    \lim_{r \to r_{H}} \frac{df(r)}{dr} > 0
    \ \ \ 
    \lim_{r \to r_{H}} \frac{f(r)}{r} = 0
    \nonumber
\end{equation}
Then the geodesic curvature in optical geometry in the near horizon limit becomes 
\begin{eqnarray}
	\lim_{r \to r_{H}} \kappa_{g}(r)
	& = & \lim_{r \to r_{H}} 
	      \bigg[ 
	         \frac{1}{\sqrt{f(r) \cdot g(r)}} 
	         \bigg( \frac{f(r)}{r} - \frac{1}{2} \frac{df(r)}{dr} \bigg) 
	      \bigg] \nonumber
	      \\
	& = & \lim_{r \to r_{H}} 
	      \bigg[ 
	         \frac{1}{\sqrt{f(r) \cdot g(r)}} 
	         \bigg( 0 - \frac{1}{2} \frac{df(r)}{dr} \bigg) 
	      \bigg] \nonumber
	      \\
	& < & 0
\end{eqnarray}
Similar to the black hole spacetime, here we have assumed the factor $\frac{1}{\sqrt{f(r) \cdot g(r)}}$ to be non-singular in the near horizon limit (such that $f(r) \cdot g(r) > 0$ is always held outside the event horizon).
In conclusion, for regular spacetime with event horizons, the geodesic curvature satisfies $\lim_{r \to r_{H}} \kappa_{g}(r) < 0$ in the near horizon limit and $\lim_{r \to \infty} \kappa_{g}(r) = 0^{+}$ at the infinite distance limit (the latter has been discussed in subsection \ref{section3a}), which is the same as the first case of figure \ref{figure1}. Under these circumstances, the existence of solutions in equation $\kappa_{g}(r)=0$ is guaranteed, providing the explicit substantiation for the presence of photon spheres in such regular spacetimes. Furthermore, assuming the continuity of geodesic curvature, it can be clearly demonstrated that the total number of photon spheres around regular spacetimes (with event horizon) must be an odd number $n = n_{\text{stable}} + n_{\text{unstable}} = 2k+1$ (where $k \in \mathbb{N}$ is a natural number).

\textbf{Regular Spacetime (without the presence of event horizon):} 
Opposite to the previous part (where the regular spacetime has an event horizon), for the regular spacetime without horizons and singularities simultaneously, the photon sphere may emerge in the entire region of spacetime $r \geq 0$. It is necessary to analyze the behavior of geodesic curvature in the center limit $r \to 0$.

For the regular spacetime without a horizon, the metric components $f(r)$ and $g(r)$ must be regular and do not change their sign, namely
\begin{align}
	f(r)=\text{finite}>0 & \ \ \ \text{when} \ r \geq 0 \nonumber
	\\
	g(r)=\text{finite}>0 & \ \ \ \text{when} \ r \geq 0 \nonumber
\end{align} 
which suggests $\lim_{r \to 0} \frac{f(r)}{r} = +\infty$ in the center limit. Meanwhile, regular spacetimes do not admit any spacetime singularities, so the derivative of the metric component must be finite in the entire spacetime
\begin{align}
	\frac{df(r)}{dr} = \text{finite} & \ \ \ \text{when} \ r \geq 0 \nonumber
\end{align}
Therefore, geodesic curvature in the center limit is reduced to 
\begin{eqnarray}
	\lim_{r \to 0} \kappa_{g}(r) 
	& = & \lim_{r \to 0} 
	      \bigg[ 
	         \frac{1}{\sqrt{f(r) \cdot g(r)}} 
	      \bigg( \frac{f(r)}{r} - \frac{1}{2} \frac{df(r)}{dr} \bigg) 
	      \bigg] \nonumber
	      \\
    & = & \lim_{r \to 0} 
    \bigg[ 
    \frac{1}{\sqrt{f(r) \cdot g(r)}}
    \bigg( +\infty - \text{finite} \bigg) \nonumber
    \\
    & = & + \infty 
\end{eqnarray}
Combining the geodesic curvature $\lim_{r \to \infty} \kappa_{g}(r) = 0^{+}$ at the infinite distance limit and $\lim_{r \to 0} \kappa_{g}(r) = + \infty$ in the center limit, this scenario is precisely what we depicted in the second case of figure \ref{figure1}, which implies that the equation $\kappa_{g}(r)=0$ possesses either no solution or even number of solutions. Consequently, the existence of photon spheres is not necessary for regular spacetimes without event horizons. And the total number of photon spheres is characterized by $n = n_{\text{stable}} + n_{\text{unstable}} = 2k$, where $k \in \mathbb{N}$ is a natural number.

\textbf{Naked Singularity Spacetime (with Finite First Order Derivatives of Metric):} The most extraordinary characteristic of naked singularity spacetime is the emergence of naked singularity that can be seen by a distant observer, which is not enclosed by event horizons. For spherically symmetric naked singularity spacetimes discussed in this work, we restrict the naked singularity in the center position $r=0$. We first consider the situation where the first-order derivatives of metric components are finite in the entire spacetime, which enables the proper definition of Riemannian curvature tensors at spacetime singularity position $r=0$ (although the geometric invariants constructed using Riemannian curvature tensors must be divergent at spacetime singularity). 

The naked singularity spacetime does not have event horizons, thus the signs of metric components $f(r)$ and $g(r)$ are kept unchanged
\begin{align}
	f(r)=\text{finite}>0 & \ \ \ \text{when} \ r \geq 0 \nonumber
	\\
	g(r)=\text{finite}>0 & \ \ \ \text{when} \ r \geq 0 \nonumber
\end{align} 
which suggests $\lim_{r \to 0} \frac{f(r)}{r} = +\infty$ in the center limit. Together with the finite first-order derivatives of the metric function
\begin{align}
	\frac{df(r)}{dr} = \text{finite} & \ \ \ \text{when} \ r >0 \nonumber
\end{align}
hence the geodesic curvature $\kappa_{g}(r)$ in the center limit is 
\begin{eqnarray}
	\lim_{r \to 0} \kappa_{g}(r) 
	& = & \lim_{r \to 0} 
	\bigg[ 
	\frac{1}{\sqrt{f(r) \cdot g(r)}} 
	\bigg( \frac{f(r)}{r} - \frac{1}{2} \frac{df(r)}{dr} \bigg) 
	\bigg] \nonumber
	\\
	& = & \lim_{r \to 0} 
	\bigg[ 
	\frac{1}{\sqrt{f(r) \cdot g(r)}}
	\bigg( +\infty - \text{finite} \bigg) \nonumber
	\\
	& = & + \infty 
\end{eqnarray}
Therefore, for naked singularity spacetime with finite first-order metric derivatives, the geodesic curvature satisfies $\lim_{r \to \infty} \kappa_{g}(r) = 0^{+}$ at the infinite distance limit and $\lim_{r \to 0} \kappa_{g}(r) = + \infty$ in the center limit, which is precisely what we have depicted in the second case of figure \ref{figure1}, suggesting $\kappa_{g}(r)=0$ possesses either no solution or even number of solutions. 
In such naked singularity spacetimes, the existence of photon spheres is not necessary, and the total number of photon spheres must be even $n = n_{\text{stable}} + n_{\text{unstable}} = 2k$, where $k \in \mathbb{N}$ is a natural number.

\textbf{Naked Singularity Spacetime (with Divergent First Order Derivatives of Metric):} Now we consider an alternative situation for naked singularity spacetime, in which the first-order derivatives of metric components become divergent at the spacetime singularity $r=0$, such that the proper definition of Riemannian curvature tensors is impossible (directly leads to the emergence of spacetime singularity).  

Because of the absence of event horizons in naked singularity spacetime, the signs of metric components $f(r)$ and $g(r)$ must be unchanged
\begin{align}
	f(r)=\text{finite}>0 & \ \ \ \text{when} \ r >0 \nonumber
	\\
	g(r)=\text{finite}>0 & \ \ \ \text{when} \ r >0 \nonumber
\end{align} 
which suggests $\lim_{r \to 0} \frac{f(r)}{r} = +\infty$ in the center limit. However, the behavior of geodesic curvature in 2-dimensional optical geometry seems subtle due to the divergent nature of the second term in brackets (which is the metric derivative $\frac{1}{2} \frac{df(r)}{dr}$, see equation (\ref{geodesic cuurvature expression})). It is necessary to discuss how the metric derivative $\frac{df(r)}{dr}$ is divergent at the spacetime singularity $r=0$. 

On the one hand, assuming the metric derivative diverges to positive infinity at spacetime singularity ($\lim_{r \to 0} \frac{df(r)}{dr} = + \infty$), the geodesic curvature $\kappa_{g}(r)$ in the center limit reduces to
\begin{eqnarray}
	& & \lim_{r \to 0} \frac{df(r)}{dr} = + \infty \nonumber
	\\
	& \Rightarrow &
	\lim_{r \to 0} \kappa_{g}(r) 
	= \lim_{r \to 0} 
	\bigg[ 
	\frac{1}{\sqrt{f(r) \cdot g(r)}} 
	\bigg( \frac{f(r)}{r} - \frac{1}{2} \frac{df(r)}{dr} \bigg) 
	\bigg] \nonumber
	\\
	& & \ \ \ \ \ \ \ \ \ \ \ \ \ 
	= \lim_{r \to 0} 
	\bigg[ 
	\frac{1}{\sqrt{f(r) \cdot g(r)}}
	\bigg( +\infty - (+\infty) \bigg) \nonumber
	\\
	& & \ \ \ \ \ \ \ \ \ \ \ \ \ 
	= \text{indefinite} 
\end{eqnarray}
In such cases, it is impossible to determine whether the equation $\kappa_{g}(r)=0$ has solutions or not, as illustrated in the third case of figure \ref{figure1}. Consequently, the existence of photon spheres in this kind of naked singularity spacetime is non-deterministic. 

On the other hand, assuming the metric derivative diverges to negative infinity at spacetime singularity ($\lim_{r \to 0} \frac{df(r)}{dr} = - \infty$), the geodesic curvature $\kappa_{g}(r)$ in the center limit becomes
\begin{eqnarray}
	& & \lim_{r \to 0} \frac{df(r)}{dr} = - \infty \nonumber
	\\
	& \Rightarrow &
	\lim_{r \to 0} \kappa_{g}(r) 
	= \lim_{r \to 0} 
	\bigg[ 
	\frac{1}{\sqrt{f(r) \cdot g(r)}} 
	\bigg( \frac{f(r)}{r} - \frac{1}{2} \frac{df(r)}{dr} \bigg) 
	\bigg] \nonumber
	\\
	& & \ \ \ \ \ \ \ \ \ \ \ \ \ 
	= \lim_{r \to 0} 
	\bigg[ 
	\frac{1}{\sqrt{f(r) \cdot g(r)}}
	\bigg( +\infty - (-\infty) \bigg) 
	\bigg] \nonumber
	\\
	& & \ \ \ \ \ \ \ \ \ \ \ \ \ 
	= +\infty
\end{eqnarray}
Combing the geodesic curvature $\lim_{r \to \infty} \kappa_{g}(r) = 0^{+}$ in the infinite distance limit and $\lim_{r \to 0} \kappa_{g}(r) = +\infty$ in the center limit, the equation $\kappa_{g}(r)=0$ possesses either no solution or even number of solutions, which is precisely what we have depicted in the second case of figure \ref{figure1}. Consequently, this kind of naked singularity spacetime (with metric derivative $\frac{df(r)}{dr}$ diverges to positive infinity at spacetime singularity) admits an even number of photon spheres.

In conclusion, for naked singularity spacetime with divergent first-order derivatives of metric, the divergent behavior of metric derivative at spacetime singularity $r=0$ greatly affects the existence of photon spheres. If the metric derivative $\frac{df(r)}{dr}$ diverges to positive infinity at spacetime singularity, the existence of photon spheres and total number of photon spheres in naked singularity spacetime are non-deterministic. Conversely, if the metric derivative $\frac{df(r)}{dr}$ diverges to negative infinity at spacetime singularity, the photon spheres do not necessarily exist in such naked singularity spacetimes, and the total number of photon spheres must be even $n = n_{\text{stable}} + n_{\text{unstable}} = 2k$ (where $k \in \mathbb{N}$ is a natural number). 

% \subsection{A Brief Summary on the Existence and Total Number of Photon Spheres}

\section{Distribution of Stable and Unstable Photon Spheres \label{section4}}

In this section, we present discussions on the distribution properties of stable and unstable photon spheres in different categories of spacetimes. The subtraction of stable and unstable photon spheres $w = n_{\text{stable}} - n_{\text{unstable}}$ is mostly focused in this section. It is one of the significantly important topics in gravity theories, through which the underlying physical properties of gravitational sources are revealed and tested. Notably, a number of recent studies have shown that the distribution of stable and unstable photon spheres can reflect the topological properties of gravitational spacetimes, and the subtraction of unstable photon spheres and stable photon spheres $w = n_{\text{stable}} - n_{\text{unstable}}$ act as a topological invariant / topological charge of spacetime \cite{Cunha2017,Cunha2018,Cunha2020,WeiSW2020}. In this work, instead of using the topological approach, we follow the geometric analysis in reference \cite{QiaoCK2024} and give a study based on curvatures in optical geometry.

\begin{figure}
	\includegraphics[width=0.51\textwidth]{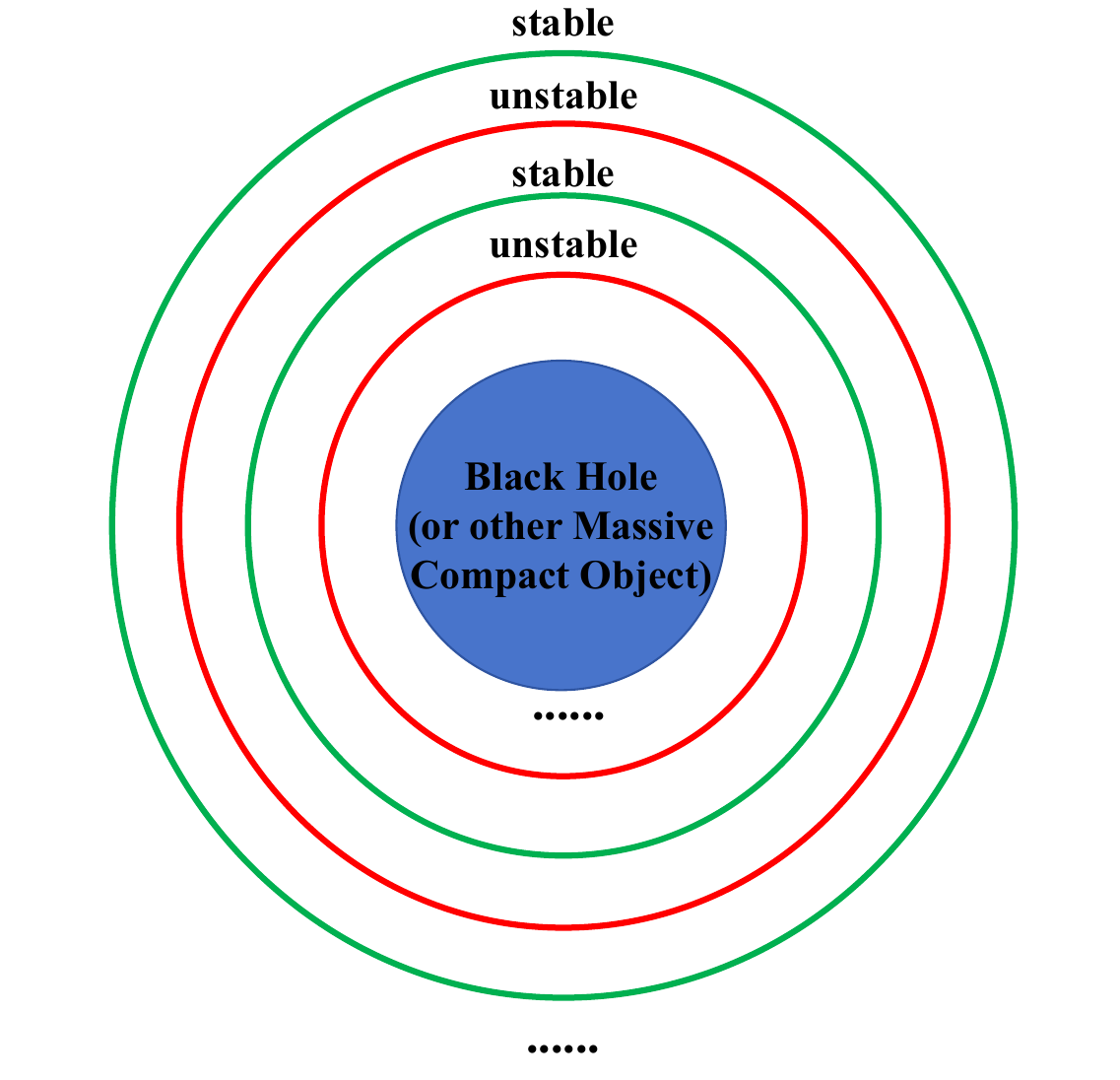}
	\caption{The stable and unstable photon spheres in spacetimes are one-to-one alternatively separated from each other.}
	\label{figure2}
\end{figure}

The most important property of photon sphere distribution is that the stable and unstable photon spheres follow a one-to-one alternatively separated distribution: 
\begin{quote}
    The stable and unstable photon spheres in spacetimes are one-to-one alternatively separated from each other, such that each unstable photon sphere is sandwiched between two stable photon spheres (and each stable photon sphere is sandwiched between two unstable photon spheres).  
\end{quote}
This distribution characteristic for stable and unstable photon spheres is illustrated in figure \ref{figure2}. Naively speaking, the one-to-one alternatively separated distribution can be understood as the successive appearance of the local maximum and local minimum of effective potentials during the photon motions (where the continuous of effective potential is assumed in literature for a number of gravitational systems). However, this feature is also closely connected with the geometric properties of optical geometry, and it can be proved rigorously by mathematical methods. The rigorous demonstration of this one-to-one alternatively separated distribution property can be accomplished through the Gauss-Bonnet theorem, which was first proposed in our recent work \cite{QiaoCK2024}. Although the original proof in reference \cite{QiaoCK2024} was proposed for black hole spacetimes, the demonstration process is irrelevant to the information on spacetime singularities and event horizons, hence it can be applied to other categories of spacetimes (such as ultra-compact object's spacetimes, regular spacetimes, and naked singularity spacetimes) without any changes. For the reader's convenience, the detailed demonstration process is briefly reviewed in appendix \ref{appendix1}. 

The one-to-one alternatively separated distribution property of stable and unstable photon spheres can easily lead to some nontrivial conclusions, especially constraining the subtraction of stable and unstable photon spheres $w = n_{\text{stable}} - n_{\text{unstable}}$ for different spacetimes and gravitational sources. For ultra compact object's spacetimes, regular spacetimes (without horizons), and naked singularity spacetimes with finite first-order derivatives (these spacetimes have an even number of photon spheres $n = n_{\text{stable}} + n_{\text{unstable}} = 2k$), the one-to-one alternatively separated distribution forces the numbers of stable and unstable photon sphere equal to each other. In these spacetimes, the subtraction of unstable and stable photon spheres $w = n_{\text{stable}} - n_{\text{unstable}}$, which serves as a topological charge to classify different kinds of spacetime, is vanished for these spacetimes. This successfully recovers a recent theorem raised by Cunha \emph{et al.} \cite{Cunha2020}, which addressed that the topological charge for ultra-compact spacetimes is $w = n_{\text{stable}} - n_{\text{unstable}} = 0$. From our geometric analysis, it is revealed that this relation is not only satisfied for ultra-compact object's spacetimes, but also holds for large categories of spacetimes with even numbers of photon spheres, including regular spacetimes without horizons, naked singularity spacetimes with finite first-order derivative, naked singularity spacetimes with first-order metric derivative $\frac{df(r)}{dr}$ divergent to negative infinity at spacetime singularity ($\lim_{r \to 0} \frac{df(r)}{dr} = - \infty$). The most common characteristic of these spacetimes is the absence of event horizons, in agreement with the Cunha's conclusion in references \cite{Cunha2020,Cunha2022}, which suggests the presence and absence of event horizons greatly influenced the topological charge of spacetimes. 

On the other hand, for spacetimes admit event horizons, such as black hole spacetimes and regular spacetime (with event horizons), the total number of photon spheres is an odd number $n = n_{\text{stable}} + n_{\text{unstable}} = 2k+1$. The one-to-one alternatively separated distribution forces the subtraction of stable and unstable photon sphere to be $w = n_{\text{stable}} - n_{\text{unstable}} = \pm 1$. To further constrain the subtraction of stable and unstable photon spheres to be $w = n_{\text{stable}} - n_{\text{unstable}} = -1$, it is necessary to demonstrate that the innermost (or outermost) photon sphere is an unstable photon sphere, such that unstable photon spheres in spacetime cannot be less than stable photon spheres. Conversely, to constrain the subtraction of stable and unstable photon spheres to be $w = n_{\text{stable}} - n_{\text{unstable}} = +1$, it is necessary to show that the innermost (or outermost) photon sphere is a stable photon sphere. Fortunately, our recent work shows that the innermost photon sphere in black hole spacetimes (which is closest to the event horizon) must be an unstable photon sphere \cite{QiaoCK2024}. In principle, the analysis is valid for any spherically symmetric spacetime with event horizons (such as regular spacetime with event horizons), not only restricted to black hole spacetimes. The detailed analyzing processes have been reviewed in appendix \ref{appendix2} for the reader's convenience. 
In this way, for black hole spacetimes and regular spacetime with event horizons, the subtraction of unstable and stable photon sphere (serves as a topological charge of spacetime) must be $w = n_{\text{stable}} - n_{\text{unstable}} = -1$, rather than $w = n_{\text{stable}} - n_{\text{unstable}} = +1$. This result successfully recovers a recent theorem on topological charge in black hole spacetimes \cite{Cunha2020,Cunha2022,WeiSW2020}.

However, for naked singularity spacetimes with first-order metric derivative $\frac{df(r)}{dr}$ divergent to positive infinity at spacetime singularity ($\lim_{r \to 0} \frac{df(r)}{dr} = + \infty$), the existence and total number of photon spheres are both non-deterministic, then the one-to-one alternatively separated distribution of stable and unstable photon spheres cannot constrain the subtraction of stable and unstable photon sphere $w = n_{\text{stable}} - n_{\text{unstable}}$ to be a definite value. In such situations, more information (such as the detailed divergent behavior of metric derivative at spacetime singularity) is needed to specify the existence and number of photon spheres.

\section{Possible Observational Method to Constrain the Photon Sphere Predictions \label{section5}}

The theoretical predictions on photon sphere / circular photon orbits might be constrained and tested from astrophysical observation data, such as very long baseline interferometry (VLBI) measurements (e.g., the optical images captured by EHT) and gravitational wave observations.

The unstable photon spheres, when viewed from the infinity observer, produce significant observational signatures in optical images of gravitational sources. Particularly, the extreme light bending near photon spheres would produce the bright emission ring in the optical images captured by VLBI observations, and they are usually called the ``photon ring'' \cite{Gralla2019,Gralla2020,Johnson2019,Chael2021,Broderick2022}. Particularly, for the black hole with an event horizon, an additional dark ``black hole shadow'' appears inside the bright photon ring, as what we have observed for M87 and Sgr*A given by EHT \cite{EHT2019a,EHT2019b,ETH2022}. Although different gravitational sources (black holes, ultra-compact objects as black hole mimickers, and naked singularities) could lead to differences in the shape and structure of optical images (as well as the lensed photon ring), the successful observation of this bright photon emission ring would inevitably give a strong evident of the existence of unstable photon spheres. The observed projected diameter of the photon emission ring, which contains radiation primarily from the lensed photon ring, is proportional to the unstable photon sphere radius $r_{\text{unstable}}$, together with influences from astrophysical accretion flows and emission models \cite{Gralla2020,Chael2021}.

When the black hole or other ultra-compact object has multiple unstable photon spheres, it could lead to notable observational signatures. For instance, the hairy black hole spacetime possessing multiple unstable photon spheres could produce more bright photon emission rings in the optical image, as well as the appearance of additional internal substructures between different bright emission rings \cite{GanQY2021,GanQY2021b}. Similar multi-ring structures can also produced by horizonless
compact objects \cite{Olmo2023}. These conclusions indicate that the presence of multiple photon ring structures (and substructures) seems to be the prominent characteristic of gravitational systems that admit multiple unstable photon spheres. 
%Moreover, the additional substructures in optical image may provide possibilities to probe the different gravitational sources with multiple photon spheres, which are proposed either in Einstein's gravity theory or in a number of alternative gravity theories. 
Although the current precision level of VLBI observation (such as EHT) may not be sufficient to observe and resolve such multiple photon emission rings and other additional substructures between photon rings, the next-generation VLBI equipment and infrastructure may have the ability to detect these structures in optical images, allowing us to discovery gravitational systems with multiple unstable photon spheres \cite{Tiede2022,Lupsasca2024}.

On the other hand, the stable photon spheres, if they exist near ultra-compact objects without event horizons, could lead to the accumulating enhancement of the gravitational wave during the gravitational perturbations, producing long-lived quasi-normal modes \cite{Cardoso2014}. These long-lived quasi-normal modes affect the ringdown process and they may have observational signatures in future gravitational wave observations. Furthermore, recent studies also pointed out that the existence of stable photon spheres (or stable light rings) and long-lived quasi-normal modes may trigger nonlinear instabilities of spacetimes, which eventually leads to the transition or collapse of such ultra-compact objects into other compact object configurations and black hole configurations in the astrophysical timescale \cite{Keir2016,Cunha2023}.

\section{Conclusions and Prospects \label{section6}}

\setlength{\tabcolsep}{2mm} % 设置表格列间距为2mm
\begin{table*}
	\caption{Features of photon spheres in different categories of spacetimes, which are obtained using our geometric analysis.}
	\label{table2}
	\vspace{2mm}
	\begin{ruledtabular}
		%\footnotesize
		\begin{tabular}{ccccll}
			Category of  & Existence of & \multicolumn{2}{c}{Distribution of Photon Spheres}  &  \multicolumn{2}{c}{Unstable and Stable}
			\\
			Spacetime & Photon Spheres & $n = n_{\text{stable}} + n_{\text{unstable}}$ &  $w = n_{\text{stable}} - n_{\text{unstable}}$ & \multicolumn{2}{c}{Photon Sphere Number}
			\\
			\hline
			\\ [-7pt]
			Black Hole         & Must Be         & $n = 2k+1$ & $w = -1$ & $n_{\text{unstable}} = k+1$ & $n_{\text{stable}} = k$ 
			\\
			Spacetime          & Existed         &
			\\
			\hline
			\\ [-7pt]
			Ultra-Compact      & Not Necessarily & $n = 2k$   & $w = 0$  & $n_{\text{unstable}} = k$ & $n_{\text{stable}} = k$ 
			\\
			Object's Spacetime & Existed     
			\\
			\hline
			\\ [-7pt]
			Regular Spacetime  & Must Be         & $n = 2k+1$ & $w = -1$ & $n_{\text{unstable}} = k+1$ & $n_{\text{stable}} = k$
			\\
			(With Horizon)     & Existed         &
			\\
			\hline
			\\ [-7pt]
			Regular Spacetime  & Not Necessarily & $n = 2k$   & $w = 0$  & $n_{\text{unstable}} = k$ & $n_{\text{stable}} = k$
			\\
			(Without Horizon)  & Existed         &
			\\
			\hline
			\\ [-7pt]
			Naked Singularity  & Not Necessarily &
			\\
			Spacetime (with    & Existed         & $n = 2k$   & $w = 0$  & $n_{\text{unstable}} = k$ & $n_{\text{stable}} = k$
			\\
			Finite $\frac{df(r)}{dr}$) & 
			\\
			\hline
			\\ [-7pt]
			Naked Singularity  & 
			\\
			Spacetime (with    & Non-deterministic & Non-deterministic & Non-deterministic & \multicolumn{2}{c}{Non-deterministic}
			\\
			$\lim_{r \to 0} \frac{df(r)}{dr} = + \infty$) & 
			\\
			\hline
			\\ [-7pt]
			Naked Singularity & Not Necessarily &
			\\
			Spacetime (with   & Existed         & $n = 2k$ & $w = 0$ & $n_{\text{unstable}} = k$ & $n_{\text{stable}} = k$
			\\
			$\lim_{r \to 0} \frac{df(r)}{dr} = - \infty$) & 
			\\
		\end{tabular}
	\end{ruledtabular}
\end{table*}

The photon sphere is a significantly important topic in the studies of black holes and other astrophysical objects. It is not only closely connected with astrophysical observations, but also unveils the underlying physics of various gravitational sources. The number and distributions of photon spheres are strongly influenced by the  features of gravitational fields, physical properties of gravitational sources, geometric and topological properties of spacetimes. In different categories of spacetimes (such as black hole spacetime, ultra-compact object's spacetimes, regular spacetimes, naked singularity spacetimes), the existence, number and distribution of photon spheres could exhibit entirely different features.

In this study, we present a general discussion on photon spheres for different categories of spacetimes, based on geometric curvatures and geometric properties of optical geometry.  
Through a geometric analysis, we successfully derive conclusions on the existence of photon spheres, total number of photon spheres $n = n_{\text{stable}} + n_{\text{unstable}}$, the subtraction of stable and unstable photon sphere numbers $w = n_{\text{stable}} - n_{\text{unstable}}$ (which can be viewed as a topological charge to clarify different categories of spacetime) in black hole spacetimes, ultra-compact objects' spacetimes, regular spacetime (with the presence and absence of event horizons), naked singularity spacetimes (with finite or divergent first-order derivative of metric). The detailed conclusions on photon spheres obtained in the present work are summarized in table \ref{table2}. 

Assuming the most general asymptotic behaviors of spacetimes (asymptotically flat, asymptotically de Sitter, and asymptotically anti de Sitter), the photon spheres in different categories of spacetimes can be summarized: 
\begin{itemize}
	\item \textbf{Black Hole Spacetime:} At least one photon sphere is existed in black hole spacetime (outside the event horizon). The total number of photon spheres in black hole spacetime is an odd number $n = n_{\text{stable}} + n_{\text{unstable}} = 2k+1$, and the subtraction of stable and unstable photon spheres fulfills $w = n_{\text{stable}} - n_{\text{unstable}} = -1$.
	
	\item \textbf{Ultra-Compact Object's Spacetime:} The photon sphere does not necessarily exist in ultra-compact object's spacetime. If photon spheres exist near the ultra-compact object's spacetime, their total number must be an even number $n = n_{\text{stable}} + n_{\text{unstable}} = 2k$. The subtraction of stable and unstable photon spheres satisfies $w = n_{\text{stable}} - n_{\text{unstable}} = 0$.
	
	\item \textbf{Regular Spacetime (With Horizon):} The photon spheres in regular spacetime (with the presence of horizons) are similar to those in black hole spacetime. There is at least one photon sphere existing in regular spacetime (outside the event horizon). The total number of photon spheres is an odd number $n = n_{\text{stable}} + n_{\text{unstable}} = 2k+1$, and the subtraction of stable and unstable photon spheres must be $w = n_{\text{stable}} - n_{\text{unstable}} = -1$. 
	
	\item \textbf{Regular Spacetime (Without Horizon):} The photon spheres in regular spacetime (without the presence of horizons) are similar to those in ultra-compact object's spacetime, where photon spheres do not necessarily exist in such spacetime. If photon spheres exist, the total number of photon spheres in regular spacetime (without horizons) must be an even number $n = n_{\text{stable}} + n_{\text{unstable}} = 2k$, and the subtraction of stable and unstable photon spheres is $w = n_{\text{stable}} - n_{\text{unstable}} = 0$.
	
	\item \textbf{Naked Singularity Spacetime (with Finite First-Order Metric Derivative):} Assuming all first-order metric derivatives are finite at the spacetime singularity $r=0$, photon spheres do not necessarily exist in this kind of naked singularity spacetime. If photon spheres exist, their total number must be even $n = n_{\text{stable}} + n_{\text{unstable}} = 2k$, and the subtraction of stable and unstable photon spheres is $w = n_{\text{stable}} - n_{\text{unstable}} = 0$.
	
	\item \textbf{Naked Singularity Spacetime (with Divergent First-Order Metric Derivative at Spacetime Singularity)}: In this kind of spacetime, the divergent behavior of metric derivative at spacetime singularity greatly influences the existence and distribution of photon spheres. Firstly, if the metric derivative $\frac{df(r)}{dr}$ diverges to positive infinity at spacetime singularity, the existence and total number of photon spheres in naked singularity spacetime are both non-deterministic. Secondly, if the metric derivative $\frac{df(r)}{dr}$ diverges to negative infinity at spacetime singularity, photon spheres do not necessarily exist in such naked singularity spacetimes. If photon spheres exist, the total number of photon spheres must be an even number $n = n_{\text{stable}} + n_{\text{unstable}} = 2k$, and the subtraction of stable and unstable photon spheres satisfies $w = n_{\text{stable}} - n_{\text{unstable}} = 0$.
\end{itemize}

Our analysis in the present study is valid for any spherically symmetric (static) gravitational system with the general spacetime metric $ds^{2} = g_{tt} dt^{2} + g_{rr} dr^{2} + g_{\theta\theta} d\theta^{2} + g_{\phi\phi} d\phi^{2}$, regardless of the detailed expressions of metric components $g_{tt}(r) = - f(r)$, $g_{rr}(r) = g(r)$. These conclusions reflect the general geometric and topological properties in different categories of spacetimes, only relying on the most general asymptotic behaviors of spacetimes (asymptotically flat, asymptotically de Sitter, and asymptotically anti de Sitter), hence they can be safely used in any specific gravitational system. 

Furthermore, our present work has the ability to stimulate some interesting studies, broadening our understanding of particle orbits in gravitational systems. One possible direction is generalizing our analysis into circular orbits of massive particles. Particularly, the existence of innermost stable circular orbits (ISCO) and the total number of marginally stable circular orbits (MSCO) for various kinds of massive particles in spacetimes are significantly important topics in recent years, for their influences on the astrophysical accretion processed and gravitational waves. To apply a similar analysis to circular orbits of massive particles, the geodesic curvature and Gaussian curvature in the Jacobi geometry of spacetimes must be utilized. The further development of our analysis into massive particle's circular orbits (such as ISCOs and MSCOs for massive particles) deserves more detailed subsequent studies. Another possible direction is extending our geometric approach to photon spheres to the axisymmetric spacetime / rotational spacetime, since a large number of gravitational sources in our universe are spinning. The extension of our approach to axisymmetric spacetimes requires more complex definitions and calculations on curvatures (such as geodesic curvature), because the corresponding optical geometry becomes a Rander-Finsler geometry in such situations. The appendix \ref{appendix3} gives an outline of how to apply our geometric approach to photon spheres in stationary and axisymmetric spacetimes.


\appendix

\section{Demonstration of the One-to-one Alternatively Separated Distribution of Stable and Unstable Photon Spheres \label{appendix1}}

In this appendix, we present a demonstration on the one-to-one alternatively separated distribution of stable and unstable photon spheres in spherically symmetric spacetimes. This one-to-one alternatively separated distribution property of photon spheres has been illustrated in figure \ref{figure2}. The demonstration presented here was first proposed in reference \cite{QiaoCK2024} for black hole spacetimes. However, the demonstration process is irrelevant to the information from spacetime singularities or event horizons. So the same conclusion is still valid in other categories of spacetimes (such as ultra-compact object's spacetimes, regular spacetimes, and naked singularity spacetimes) without any changes.

\begin{figure}
	\includegraphics[width=0.525\textwidth]{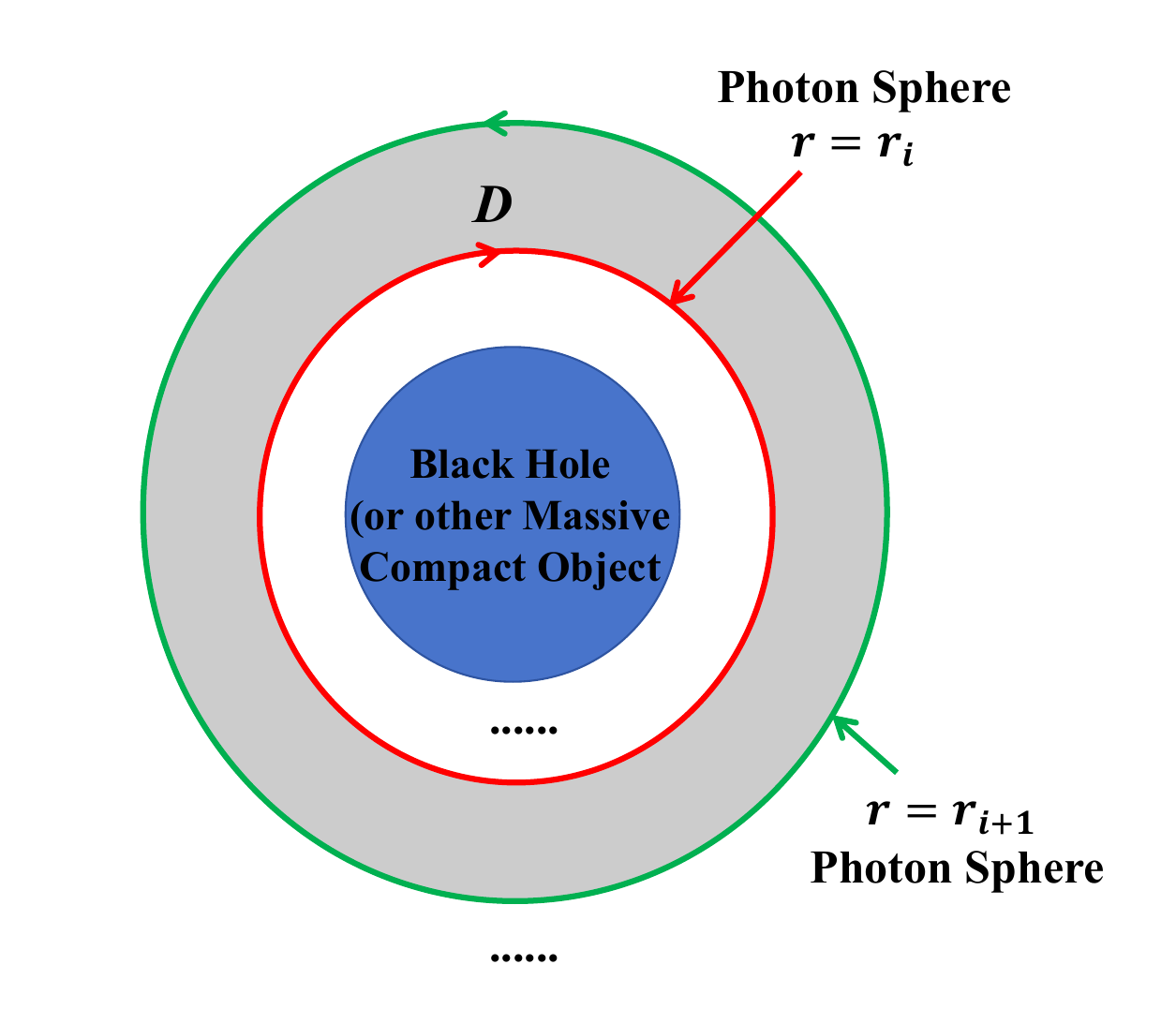}
	\caption{The choice of region $D$ in 2-dimensional optical geometry in the Gauss-Bonnet theorem, which is used to demonstrate that stable and unstable photon spheres in spacetime are one-to-one alternatively separated from each other.}
	\label{figure3}
\end{figure}

\begin{figure*}
	\includegraphics[width=0.495\textwidth]{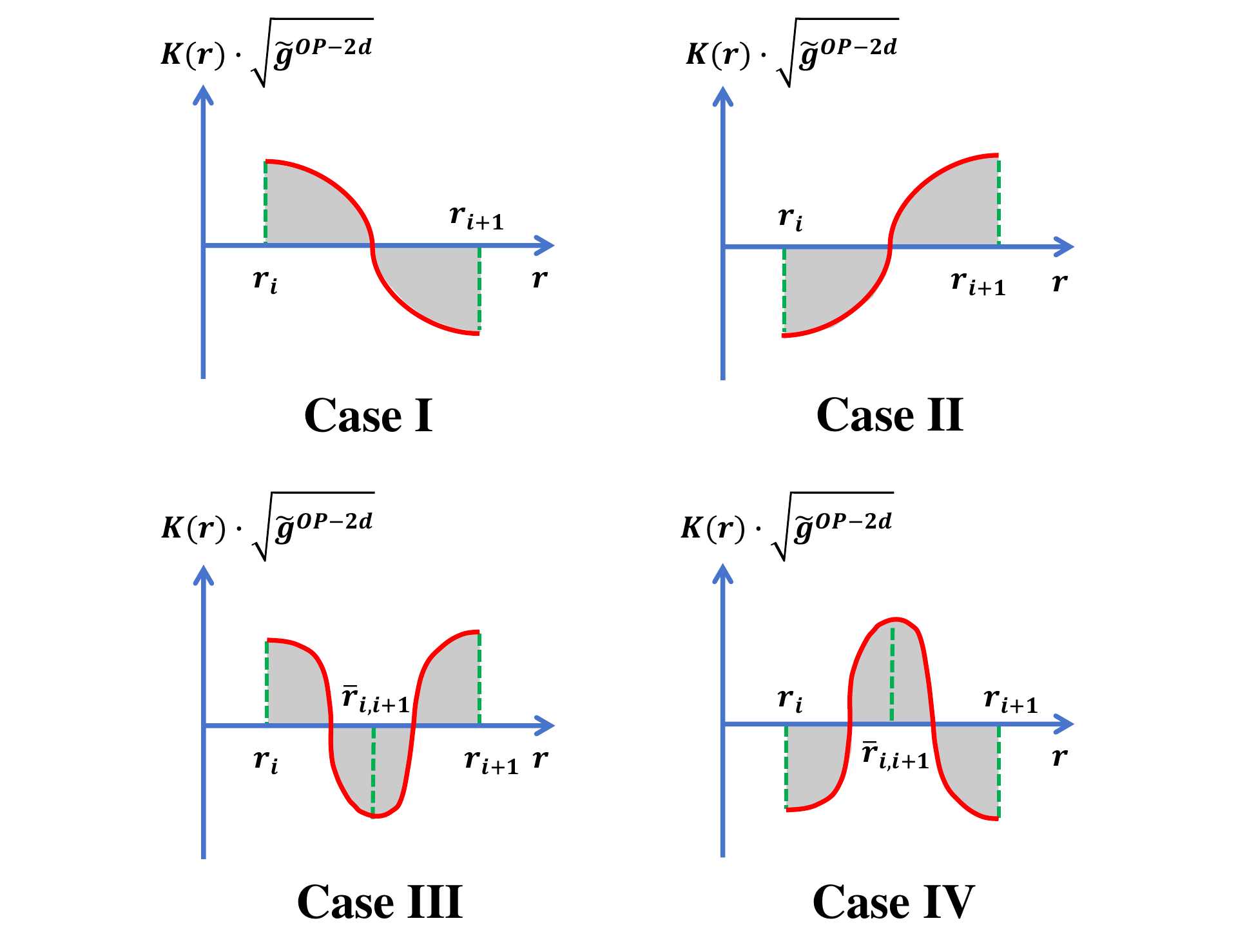}
	\includegraphics[width=0.495\textwidth]{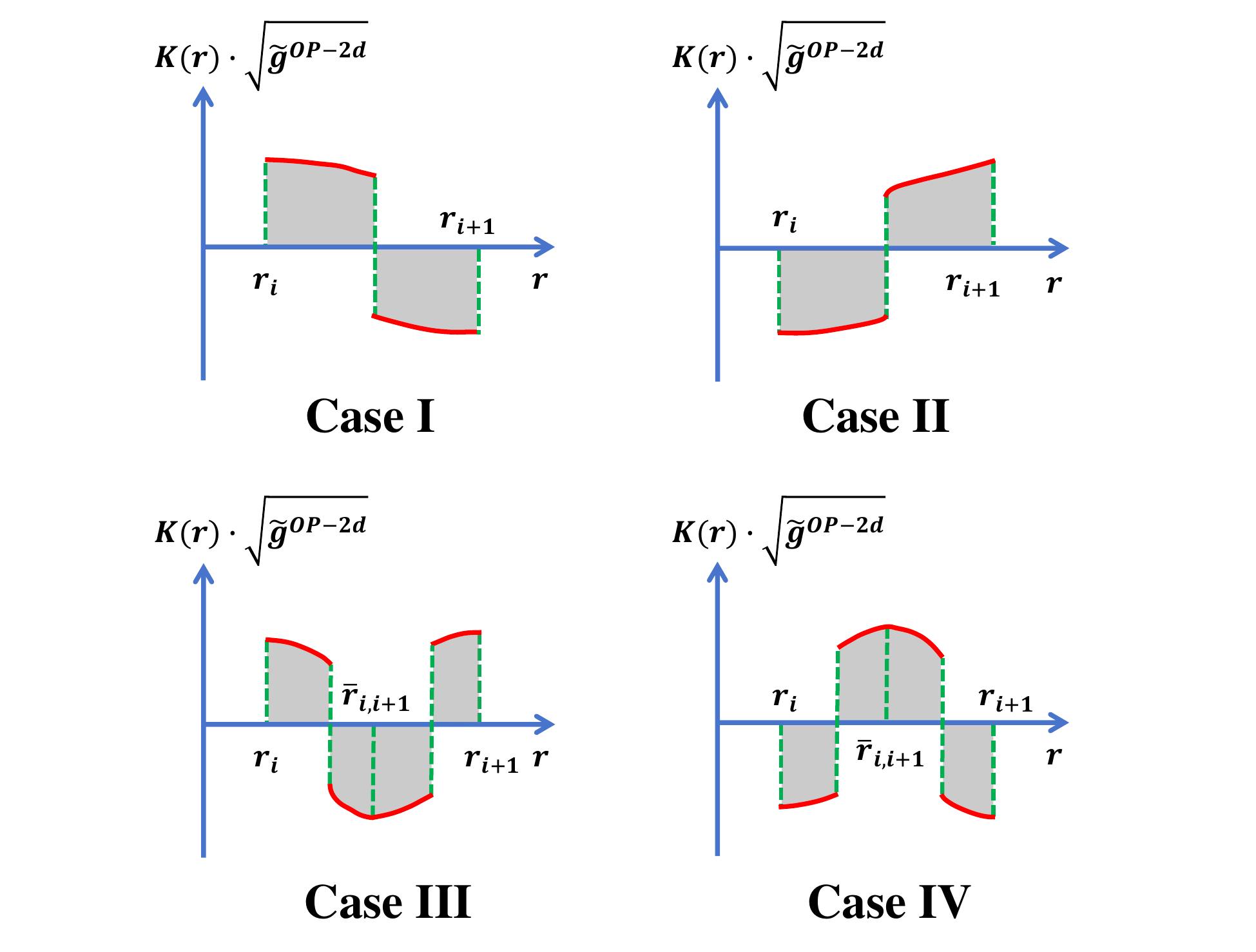}
	\caption{This figure summarizes the four different cases in which the surface integral of Gaussian curvature vanishes in region $D$ (namely $\int_{D}\mathcal{K}\cdot dS = 0$). \textbf{Case I:} the inner boundary photon sphere is stable and the outer boundary photon sphere is unstable, with $\mathcal{K}(r_{i}) > 0$ and $\mathcal{K}(r_{i+1}) < 0$. \textbf{Case II:} the inner boundary photon sphere is unstable and the outer boundary photon sphere is stable, with $\mathcal{K}(r_{i}) < 0$ and $\mathcal{K}(r_{i+1}) > 0$. \textbf{Case III:} both the inner and outer boundary photon spheres are stable, with $\mathcal{K}(r_{i}) > 0$ and $\mathcal{K}(r_{i+1}) > 0$. \textbf{Case IV:} both the inner and outer boundary photon spheres are unstable, with $\mathcal{K}(r_{i}) < 0$ and $\mathcal{K}(r_{i+1}) < 0$. In the last two cases, a further analysis suggests that the inner and outer boundary photon spheres at $r=r_{i}$ and $r=r_{i+1}$ cannot be adjacent, and an additional photon sphere must exist between $r_{i}$ and $r_{i+1}$. It is worth noting that the $\sqrt{\tilde{g}^{\text{OP-2d}}}$ in surface element is always positive, hence the sign of Gaussian curvature can be easily seen from this figure. In particular, the left panel shows the situations where Gaussian curvature is continuous in region $D$, while the right panel represents the situations where Gaussian curvature admits a finite number of discontinuous points with respect to radial coordinate $r$.}
	\label{figure4}
\end{figure*}

To demonstrate the one-to-one alternatively separated distribution property for stable and unstable photon spheres, we apply the Gauss-Bonnet theorem in 2-dimensional optical geometry of spacetime. Firstly, a region $D$ must be selected in optical geometry for the application of Gauss-Bonnet theorem. The selected region $D$ is an annular zone bounded by two adjacent photon spheres, which is shown in figure \ref{figure3}. It is a complexly connected region, and its Euler characteristic number is $\chi(D)=0$. The application of Gauss-Bonnet theorem within this region yields
\begin{equation}
	\int_{D}\mathcal{K}\cdot dS = 2 \pi \chi(D) = 0 
\end{equation}
which means the ``averaged'' Gaussian curvature between two adjacent photon spheres is zero.
In particular, from one of the authors' recent works \cite{HuangY2022,HuangY2023}, a simplification on the surface integral of Gaussian curvature in the 2-dimensional optical geometry can be carried out 
\begin{eqnarray}
	\int_{D}\mathcal{K}\cdot dS 
	& = & 
	\int_{0}^{2\pi}d\phi \int_{r_{i}}^{r_{i+1}} \mathcal{K}(r) \sqrt{\tilde{g}^{\text{OP-2d}}(r)} dr \nonumber
	\\
	& = & 2 \pi \big[ H(r_{i+1})-H(r_{i}) \big] = 0 
	\label{Gauss-Bonnet reduce}
\end{eqnarray}
where function $H(r)$ is defined as
\begin{equation}
	H(r) = - \frac{1}{2\sqrt{\tilde{g}^{\text{OP-2d}}}}
	\cdot \frac{\partial \tilde{g}^{\text{OP-2d}}_{\phi\phi}}{\partial r}
	= - \sqrt{\tilde{g}^{\text{OP-2d}}_{\phi\phi}} \cdot \kappa_{g}(r) 
	\label{function H}
\end{equation} 
The last equality sign directly comes from the geodesic curvature expression for circular curves given in equation (\ref{geodesic cuurvature expression}). Since the inner and outer boundaries of region $D$ are both photon spheres with vanishing geodesic curvature, it is necessary to have $\kappa_{g}(r_{i+1}) = \kappa_{g}(r_{i}) = 0 \Rightarrow H(r_{i+1}) = H(r_{i}) = 0$ for the boundaries of this annular region $D$.

There are four different cases that could lead to the surface integral of Gaussian curvature vanishing in region $D$, which have been summarized in figure \ref{figure4}. The vanishing of surface integral $\int_{D}\mathcal{K}\cdot dS = 2 \pi \int_{r=r_{i}}^{r=r_{i+1}} \mathcal{K}(r) \sqrt{\tilde{g}^{\text{OP-2d}}} dr = 0$ suggests the shadow areas depicted in figure \ref{figure4} in the integration are zero. 
In the first two cases, the inner and the outer photon spheres are of different types (one is a stable photon sphere, and the other is an unstable photon sphere), resulting in one-to-one alternating distributions of stable and unstable photon spheres. However, in case III and case IV, the inner and outer photon spheres are of the same types, which violates the one-to-one alternatively separated distribution of stable and unstable photon spheres in figure \ref{figure2}. Fortunately, a further analysis suggests that there must be an additional photon sphere between the inner boundary photon sphere $r=r_{i}$ and outer boundary photon sphere $r=r_{i+1}$. Take the case III as an example to show this conclusion. If the inner and outer photon spheres are both stable photon spheres with  $\mathcal{K}(r=r_{i}) > 0$ and $\mathcal{K}(r=r_{i+1}) > 0$, we can always find a position $\bar{r}_{i,i+1}$ such that the region $D$ can be divided into two parts. This division requires that integrations of Gaussian curvature $\int \mathcal{K}(r) \sqrt{\tilde{g}^{\text{OP-2d}}(r)} dr$ over intervals $[r_{i}, \bar{r}_{i,i+1}]$ and $[\bar{r}_{i,i+1}, r_{i+1}]$ are vanished (and the shadow areas in intervals $[r_{i}, \bar{r}_{i,i+1}]$ and $[\bar{r}_{i,i+1}, r_{i+1}]$ are both zero). From the simplified integration formulas in (\ref{Gauss-Bonnet reduce}) and (\ref{function H}), it is evident that this position satisfies $H(\bar{r}_{i,i+1}) = H(r_{i}) = H(r_{i+1})$ and $\kappa_{g}(\bar{r}_{i,i+1}) = \kappa_{g}(r_{i+1}) = \kappa_{g}(r_{i}) = 0$, indicating $r=\bar{r}_{i,i+1}$ must be the location of another photon sphere. Furthermore, the negative Gaussian curvature $\mathcal{K}(\bar{r}_{i,i+1}) < 0$ indicates that this new photon sphere at $r=\bar{r}_{i,i+1}$ is an unstable photon sphere, which is opposite to the inner and outer photon spheres at $r=r_{i}$, $r=r_{i+1}$. A similar analysis can be carried out for case IV, where an additional stable photon sphere must exist between the inner and outer unstable photon spheres at $r=r_{i}$ and $r=r_{i+1}$. This analysis remains valid even if there are finite number of discontinuous points in the Gaussian curvature $\mathcal{K}(r)$ (as presented in the right panel of figure \ref{figure4}) \footnote{Conventionally, the Gauss-Bonnet theorem is applied to a smooth region, where the Gaussian curvature is continuous in the entire region of $D$ (and the geodesic curvature along boundary $\partial D$ is also continuous). The discontinuity of curvatures may potentially bring up additional terms contributing to the Gauss-Bonnet theorem (which may depend on the specific behaviors of the discontinuous points). However, exploring the possible extended formulas for the Gauss-Bonnet theorem in cases with discontinuities is far beyond the scope of our work. Our analysis is carried out assuming that the classical Gauss-Bonnet theorem $\int_{D} \mathcal{K} \dot dS + \int_{\partial D} \kappa_{g} dl + \sum_i \theta_{i} = 2 \pi \chi(D)$ is still valid. \label{Gauss-Bonnet validity}}. 
In conclusion, among all four cases, we have proven that two stable photon spheres (or two unstable photon spheres) cannot be adjacent. The stable and unstable photon spheres in a spherically symmetric spacetime must be one-to-one alternatively separated from each other. Each stable photon sphere is sandwiched between two nearby unstable photon spheres, and each unstable photon sphere is sandwiched between two stable photon spheres.

\begin{figure*}
	\includegraphics[width=0.495\textwidth]{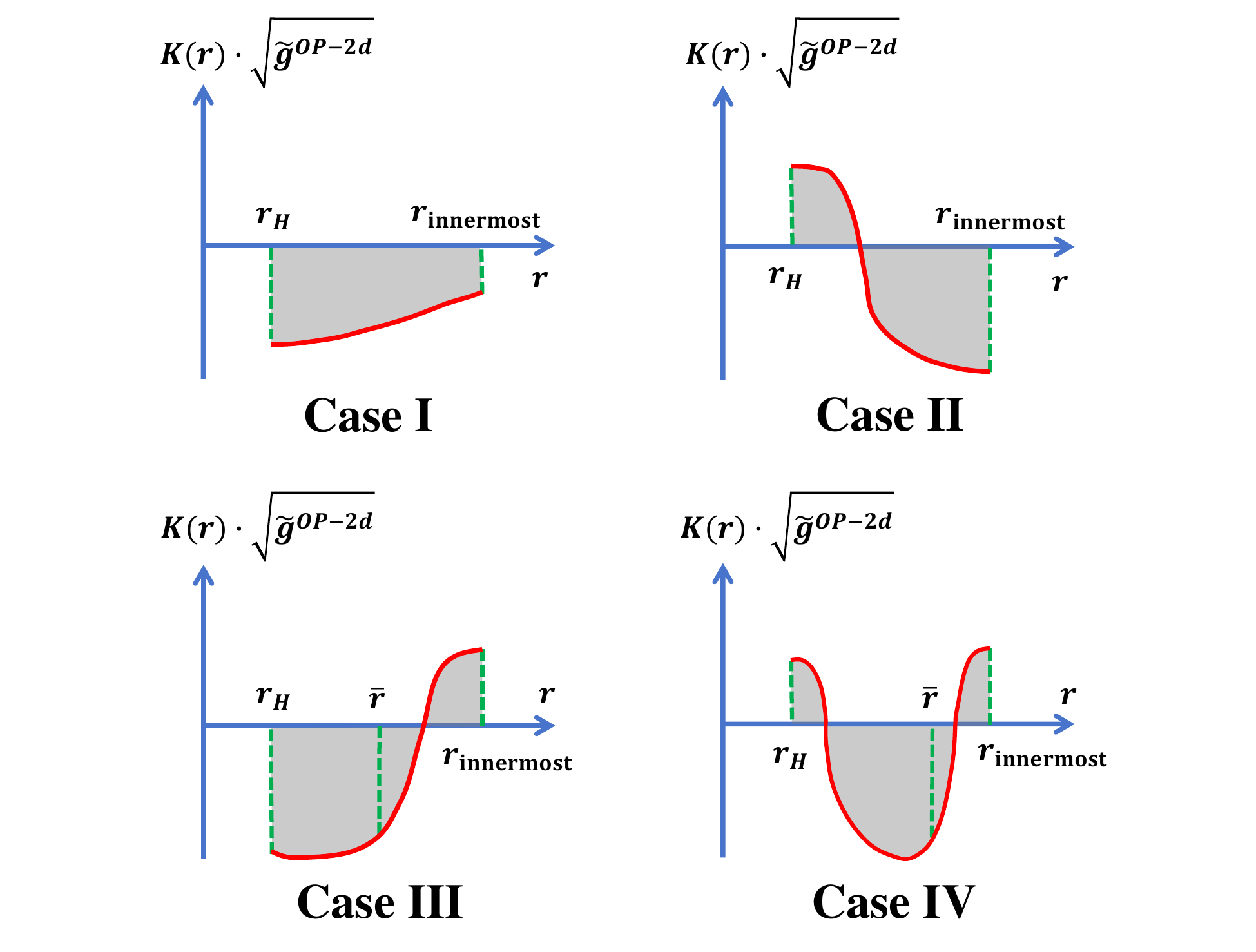}
	\includegraphics[width=0.495\textwidth]{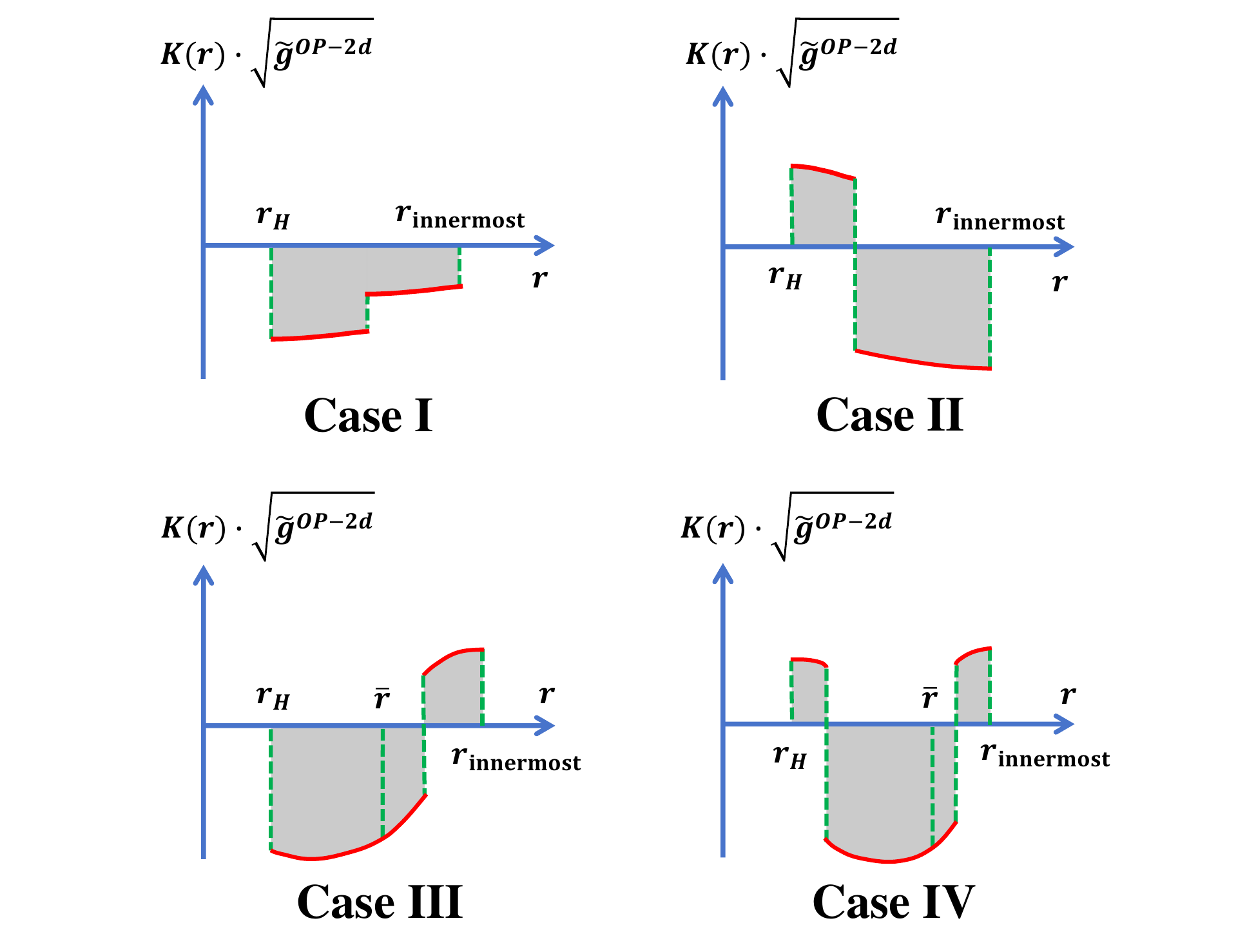}
	\caption{This figure summarizes the four different cases in which the surface integral of Gaussian curvature satisfies $\int_{D}\mathcal{K}\cdot dS < 0$. \textbf{Case I:} the innermost photon sphere is unstable, with $\mathcal{K}(r_{H}) < 0$ and $\mathcal{K}(r_{\text{innermost}}) < 0$. \textbf{Case II:} the innermost photon sphere is unstable, with $\mathcal{K}(r_{H}) > 0$ and $\mathcal{K}(r_{\text{innermost}}) < 0$. \textbf{Case III:} the innermost photon sphere is stable, with $\mathcal{K}(r_{H}) < 0$ and $\mathcal{K}(r_{\text{innermost}}) > 0$. \textbf{Case IV:} the innermost photon sphere is stable, with $\mathcal{K}(r_{H}) > 0$ and $\mathcal{K}(r_{\text{innermost}}) > 0$. In the last two cases, a further analysis suggests that an additional photon sphere $r=\bar{r}$ must exist between the event horizon $r=r_{H}$ and the faked ``innermost'' photon sphere $r=r_{\text{innermost}}$. It is worth noting that the $\sqrt{\tilde{g}^{\text{OP-2d}}}$ in surface element is always positive, hence the sign of Gaussian curvature can be easily seen from this figure. In particular, the left panel shows the situations where Gaussian curvature is continuous in region $D$, while the right panel represents the situations where Gaussian curvature admits a finite number of discontinuous points with respect to radial coordinate $r$.}
	\label{figure5}
\end{figure*}

\section{Demonstration of the Innermost Photon Sphere Near Event Horizon to be Unstable \label{appendix2}}

In this appendix, we present a detailed demonstration of the innermost photon sphere near the event horizon to be unstable, for the reader's convenience. This conclusion was first demonstrated in reference \cite{QiaoCK2024} for black hole spacetimes. However, as will be seen in the following, the whole demonstration process is irrelevant to the information inside the event horizon, therefore it can be applied to any spacetimes with event horizons. Finally, we arrive at a generalized conclusion: if the spacetime has an event horizon (e.g., black hole spacetime, regular spacetime), the innermost photon sphere closest to the event horizon must be an unstable photon sphere. 

In the demonstration, the Gauss-Bonnet theorem should be applied to a similar annular region depicted in figure \ref{figure3}. In this case, the inner boundary circle of region $D$ is approached to the event horizon $r_{\text{min}} \to r_{H}$ and the outer boundary circle is the innermost photon sphere $r_{\text{max}} = r_{\text{innermost}}$. The application of Gauss-Bonnet theorem in such region leads to
\begin{equation}
	\int_{D}\mathcal{K}\cdot dS + \int_{\partial D} \kappa_{g} \cdot dl = 2 \pi \chi(D) = 0
\end{equation}
The outer boundary of region $D$ is the innermost photon sphere, so the geodesic curvature vanishes. The inner boundary of region $D$ is approaching the event horizon $r_{H}$, and the contour integral of geodesic curvature gives 
\begin{eqnarray}
	\int_{\partial D} \kappa_{g} \cdot dl 
	& = & - \int_{C_{r}(r \to r_{H})} \kappa_{g}(r) \cdot dl \nonumber
	\\
	& = & - \int_{\phi=0}^{\phi=2\pi} \kappa_{g}(r_{H}) \cdot \sqrt{\tilde{g}_{\phi\phi}^{\text{OP-2d}}(r_{H})} \cdot d\phi \nonumber
	\\
	& = & - C(r_{H}) \cdot \lim_{r \to r_{H}} \kappa_{g}(r) \nonumber
	\\
	& \propto & - 2 \pi r_{H} \cdot \lim_{r \to r_{H}} \kappa_{g}(r)
\end{eqnarray}
Here, $C(r_{H}) \sim 2 \pi r_{H}$ represents the circumference of inner boundary $r \to r_{H}$ in 2-dimensional optical geometry. Note that the direction of inner boundary of region $D$ illustrated in figure \ref{figure3} is opposite to the increasing of azimuthal angle $\phi$, which contributes to an additional minus sign 
\footnote{The direction of the contour integral $\int_{\partial D}\kappa_{g} dl$ illustrated in figure (which corresponds to figure \ref{figure3}) in reference \cite{QiaoCK2024} is not correct. The correct contour integral direction is what we have shown in figure \ref{figure3}. We shall write an erratum/correction of reference \cite{QiaoCK2024} in the near future.}. 
Hence we obtain the following relation for the surface integral of Gaussian curvature
\begin{eqnarray}
	\int_{D}\mathcal{K}\cdot dS 
	& = & - \int_{\partial D} \kappa_{g} \cdot dl \nonumber
	\\ 
	& \propto & 2 \pi r_{H} \cdot \lim_{r \to r_{H}} \kappa_{g}(r) 
	< 0 
\end{eqnarray}
In the derivation, the geodesic curvature in the near horizon limit $\lim_{r \to r_{H}}\kappa_{g}(r) < 0$ obtained in section \ref{section3b} is used for black hole spacetime and regular spacetime. 

There are four different cases that could lead to the surface integral of Gaussian curvature satisfying $\int_{D}\mathcal{K}\cdot dS < 0$. These cases have been summarized in figure \ref{figure5}. The negative value of the surface integral $\int_{D}\mathcal{K}\cdot dS = 2 \pi \int_{r_{H}}^{r_{\text{innermost}}} \mathcal{K}(r) \sqrt{\tilde{g}^{\text{OP-2d}}(r)} dr < 0$ indicates that the shadow areas depicted in figure \ref{figure5} in the integration are less than zero. In the first two cases, the innermost photon spheres are both unstable photon spheres with negative Gaussian curvature. However, in the last two cases, it appears that the innermost photon spheres are stable photon spheres instead of the unstable ones as expected. Fortunately, a simple analysis suggests that an additional photon sphere must exist between the event horizon $r=r_{H}$ and the faked ``innermost'' photon sphere $r=r_{\text{innermost}}$. The truly innermost photon sphere should be located at a smaller radius $r=\bar{r}$, rather than $r=r_{\text{innermost}}$. From the case III and case IV of figure \ref{figure5}, it is always possible to find a new position $r=\bar{r}$ in the interval $[r_{H}, r_{\text{innermost}}]$, such that the integral $\int \mathcal{K}(r) \sqrt{\tilde{g}^{\text{OP-2d}}(r)} dr$ vanishes in the interval $[\bar{r}, r_{\text{innermost}}]$ (and the shadow areas in this interval are zero). Utilizing the simplified integration formulas in equations (\ref{Gauss-Bonnet reduce}) and (\ref{function H}), the vanishing of integral suggests that $H(\bar{r}) = H(r_{\text{innermost}}) = 0$ and $\kappa_{g}(\bar{r}) = \kappa_{g}(r_{\text{innermost}}) = 0$, which suggests that $r=\bar{r}$ must be the proper position for another photon sphere. Moreover, the negative Gaussian curvature constrains this new photon sphere at position $r=\bar{r}$ to be an unstable photon sphere. In summary, in all four cases, we have demonstrated the innermost photon sphere must be an unstable photon sphere. Notably, our demonstration processes presented here still hold even if the Gaussian curvature admits a finite number of discontinuous points, as illustrated in the right panel of figure \ref{figure5} 
\footnote{Similar to the discussions in appendix \ref{appendix1} and footnote \ref{Gauss-Bonnet validity}, our analysis is carried out assuming that the classical Gauss-Bonnet theorem $\int_{D} \mathcal{K} \dot dS + \int_{\partial D} \kappa_{g} dl + \sum_i \theta_{i} = 2 \pi \chi(D)$ is valid.}.

\section{Possible Extension of Our Geometric Approach to the Axisymmetric Spacetime \label{appendix3}}

The analyzing method on circular photon orbits presented in this work, which is based on geometric approach and curvatures in optical geometry, can be applied to the axisymmetric spacetime. This appendix provides an outline of the possible extension of our approach to axisymmetric spacetimes.

In the axisymmetric spacetime, the optical geometry can still be constructed through the null constraint $d\tau^{2}=-ds^{2}=0$ on a 4-dimensional spacetime geometry, the same as what we have introduced in section \ref{section2}. This construction of optical geometry of any axisymmetric spacetime would result in a Randers-Finsler geometry \cite{Werner2012,Asida2017,Jusufi2017,LiZH2019a,HuangY2023}, rather than the Riemannian optical geometry for the static and spherically symmetric spacetime cases \footnote{Particularly, recent studies proposed by O. L. Andino \emph{et. al.} show that the Randers-Finsler geometry can be bypassed through a ``generalized'' construction of the optical geometry / Jacobi geometry \cite{Andino2019,Andino2024,Andino2025}. This new construction provides insights into the optical geometry / Jacobi geometry and deserves further investigations. However, in this appendix, we follows the construction of optical geometry given by G. W. Gibbons \emph{et. al.} \cite{Gibbons2009,Gibbons2008,Werner2012}, which leads to a Randers-Finsler optical geometry for axisymmetric spacetimes.}. The optical geometry of axisymmetric spacetimes gives
\begin{equation}
	\underbrace{ds^{2} = g_{\mu\nu}dx^{\mu}dx^{\nu}}_{\text{Spacetime Geometry}}
	\ \ \overset{d\tau^{2}=-ds^{2}=0}{\Longrightarrow} \ \ 
	\underbrace{dt = \sqrt{\alpha_{ij}dx^{i}dx^{j}} + \beta_{i}dx^{i}}_{\text{Optical Geometry}}
	\label{optical geometry}
\end{equation}
with $T$ to be the tangent vector of the continuous null curve.

For any axisymmetric spacetime with the general metric form
\begin{equation}
	ds^{2} = g_{tt}dt^{2}+2g_{t\phi}dtd\phi+g_{rr}dr^{2}+g_{\theta\theta}d\theta^{2}+g_{\phi\phi}d\phi^{2}
\end{equation} 
Imposing of null constraint $d\tau^{2}=-ds^{2}=0$ in spacetime geometry gives 
\begin{eqnarray}
	dt & = & \sqrt{-\frac{g_{rr}}{g_{tt}} \cdot dr^{2} - \frac{g_{\theta\theta}}{g_{tt}} \cdot d\theta^{2} + \frac{g_{t\phi}^{2}-g_{tt}g_{\phi\phi}}{g_{tt}^{2}} \cdot d\phi^{2}} \nonumber
	\\
	&   & - \frac{g_{t\phi}}{g_{tt}}\cdot d\phi \label{Optical geometry axisymmetric}
\end{eqnarray}
Here, the arc length parameter / spatial distance parameter in optical geometry becomes the stationary time coordinate $t$. Mathematically, the optical geometry in (\ref{Optical geometry axisymmetric}) gives a Renders-Finsler manifold. The Renders-Finsler geometry is an extension of the Riemannian geometry \cite{ChernSS}, allowing the separation of arc-length / spatial distance into two parts 
\begin{equation}
	ds = \sqrt{\alpha_{ij}(x)dx^{i}dx^{j}} + \beta_{i}(x)dx^{i} . \label{Renders geometry}
\end{equation}
The first part $\alpha_{ij}$ is a Riemannian metric, and the second part $\beta=\beta_{i}dx^{i}$ quantifies the departure of this Renders-Finsler geometry from the Riemannian geometry. It is natural to see that the Renders-Finsler geometry recovers a Riemannian geometry if and only if $\beta=0$.

In the axisymmetric spacetime, the photon sphere is no longer existed in general (because the photon sphere is a spherically symmetric surface). This makes the circular photon orbits a bit more complex than those in spherically symmetric spacetimes. One specific classification of photon spheres is of great interest in recent studies --- the light rings in the equatorial plane of an axisymmetric spacetime. To analyze the existence and distribution of light rings, the optical geometry in (\ref{Optical geometry axisymmetric}) must be constrained into the equatorial plane
\begin{eqnarray}
	dt	& = & \sqrt{-\frac{g_{rr}}{g_{tt}} \cdot dr^{2} + \frac{g_{t\phi}^{2}-g_{tt}g_{\phi\phi}}{g_{tt}^{2}} \cdot d\phi^{2}} 
	- \frac{g_{t\phi}}{g_{tt}}\cdot d\phi  \nonumber
	\\ \label{Optical geometry 2D axisymmetric}
\end{eqnarray}

The light rings in any axisymmetric spacetimes are null geodesics in 4-dimensional spacetime geometry, they become spatial geodesics when transforming into optical geometry. In this way, any light rings must be geodesics in 2-dimensional Randers-Finsler geometry, and the covariant derivative for these light rings vanishes
\begin{equation}
	\text{Light Rings}
	\ \ \Leftrightarrow \ \
	D_{T}^{(F)} T = 0 
\end{equation}
with $T=\frac{dx}{dt}$ to be the tangent vector in 2-dimensional optical geometry, and $D_{T}^{(F)}$ is the covariant derivative operator along this tangent direction in the 2-dimensional Randers-Finsler geometry. 

The concept of geodesic curvature can also be generalized into Randers-Finsler geometry.
Firstly, for a continuous curve $\gamma$ in the 2-dimensional Randers-Finsler geometry, its geodesic curvature can be defined proportional to the inner product of the covariant derivative of tangent vector $D_{T}^{(F)} \tilde{T}$ and a unit vector $\tilde{V}$ \cite{Shimada2010}, through the relation 
\footnote{Here, the superscript $(F)$ denotes the inner product in the Finsler geometry. It is worth noting that the mathematical definition on the inner product of two vector fields in Finsler geometry depends not only on base point $x$, but also on reference vector $y$ (with $(x,y) \in TM$) \cite{ChernSS}. This is different from the cases in Riemannian geometry, in which the Riemannian metric depends solely on the base point $x \in M$. Therefore, in order to write the inner product $<V,W>$ in Finsler geometry appropriately, it is necessary to add a pair $(x,y)$ in the subscript to specify the reference vector. In the definition of geodesic curvature $\kappa_{g}^{(F)}$ for a continuous curve, it is convenient to choose the reference vector to be the tangent vector of this curve, namely $y=T$.}  
\begin{equation}
	\kappa_{g}^{(F)} \propto \big< D_{T}^{(F)} \tilde{T} \cdot \tilde{V} \big>^{(F)}_{(x,T)}
	\label{geodesic curvature Finsler}
\end{equation}
where $\tilde{T}=T/F(x,T)$ is the unit tangent vector of the continuous curve $\gamma$, and $\tilde{V}$ is a unit vector in the 2-dimensional Randers-Finsler geometry that is perpendicular to the tangent vector $\tilde{T}$ (that is $<\tilde{T},\tilde{V}>^{(F)}_{(x,T)}=0$). On the other hand, since the Randers-Finsler geometry is a generalization of the Riemannian geometry, it is natural to expect that the geodesic curvature of a continuous curve in 2-dimensional Randers-Finsler geometry can be expressed using the Riemannian geometry part $\alpha_{ij}dx^{i}dx^{j}$ and non-Riemannian part $\beta$ in expression (\ref{Renders geometry}). So we have the following expression of geodesic curvature
\begin{equation}
	\kappa_{g}^{(F)} = \kappa_{g}^{(\alpha)} + \kappa_{\beta}^{(\alpha)}
\end{equation}	
where $\kappa_{g}^{(\alpha)}$ is the geodesic curvature of the same continuous curve with respect to the Riemannian metric $\alpha_{ij}dx^{i}dx^{j}$, and $\kappa_{\beta}^{(\alpha)}$ is the additional contribution to $\kappa_{g}^{(\alpha)}$ due to the existence of non-Riemannian part $\beta$. 

The above definition of geodesic curvature in (\ref{geodesic curvature Finsler}) measures the departure of a continuous curve from being a geodesic curve in 2-dimensional Randers-Finsler geometry. Since light rings maintain the geodesic property when transformed into optical geometry, their geodesic curvature naturally vanished.
\begin{eqnarray}
	\text{Light Rings}
	& \Leftrightarrow & 
	D_{T}^{(F)} T = 0 \nonumber
	\\
	& \Leftrightarrow &  
	\kappa_{g}^{(F)} = \kappa_{g}^{(\alpha)} + \kappa_{\beta}^{(\alpha)} = 0
\end{eqnarray}
Furthermore, through a tedious derivation (which is not presented here), the contribution caused by non-Riemannian part $\beta$ can be calculated as \cite{Asida2017,QiaoCK2025b} 
\footnote{Particularly, in the upcoming work \cite{QiaoCK2025b}, we adopt the regularization such that an additional minus sign contribute to the $\kappa_{\beta}^{\alpha}$ for counter-rotating motion of lights (with angular velocity $\Omega = \frac{d\phi}{dt} < 0$), which leads to the following equation for light rings
\begin{equation}
	\kappa_{g}^{(\alpha)} = - \kappa_{\beta}^{(\alpha)}
	= - \frac{\text{Sign}(\Omega)}{\sqrt{\alpha_{rr}\alpha_{\phi\phi}}}
	\cdot \frac{\partial \beta_{\phi}}{\partial r} \nonumber
\end{equation} }
\begin{eqnarray}
	&                 &
	\kappa_{g}^{(F)} = \kappa_{g}^{(\alpha)} + \kappa_{\beta}^{(\alpha)} = 0 \nonumber
	\\
	& \Leftrightarrow &
	\kappa_{g}^{(\alpha)} = - \kappa_{\beta}^{(\alpha)}
	= - \frac{1}{\sqrt{\alpha_{rr}\alpha_{\phi\phi}}}
	\cdot \frac{\partial \beta_{\phi}}{\partial r} 
	%\\
	%&                 & 
	%\ \ \ \ \ \ \ \ \ \ \ \ \ \ \ \ \ \ 
	%= - \frac{1}{\sqrt{\alpha_{rr}\alpha_{\phi\phi}}}
	%\cdot \frac{\partial}{\partial r} \bigg( - \frac{g_{t\phi}}{g_{tt}} \bigg)
	%\ \ \ \ 
	\label{light ring geodesic curvature condition}
\end{eqnarray}
This is the geodesic curvature condition for light rings based on the 2-dimensional Randers-Finsler optical geometry.

The above discussions provide an outline of our geometric approach extended to axisymmetric spacetimes. Similar to the static and spherically symmetric spacetime cases, the existence of light rings in the equatorial plane of axisymmetric spacetimes suggests that the equation $\kappa_{g}^{(F)} = 0$ (or the geodesic curvature condition given in equation (\ref{light ring geodesic curvature condition})) admits at least one solution \cite{QiaoCK2024}. To provide an analysis of the existence and distribution of light rings in general axisymmetric gravitational systems, the behavior of geodesic curvature $\kappa_{g}^{(F)} = \kappa_{g}^{(\alpha)} + \kappa_{\beta}^{(\alpha)}$ for a circular curve must be analyzed in detail, especially in the center limit $r \to 0$, near-horizon limit $r \to r_{H}$, and infinite distance limit $r \to \infty$. In principle, one can carry out this approach to give a comprehensive analysis of light rings for different categories of spacetime (black hole spacetime, ultra-compact object's spacetime, regular spacetime and, naked singularity spacetime). However, this analysis is beyond the scope of the present work. An exploration of light rings for these spacetimes will be carried out in future works.

%\begin{acknowledgments}        
%\end{acknowledgments}

\begin{acknowledgments}
	This work is supported by the Natural Science Foundation of China (Grant No. 1240051542), the Scientific and Technological Research Program of Chongqing Municipal Education Commission (Grant No. KJQN202201126), the Natural Science Foundation of Chongqing Municipality (Grant No. CSTB2022NSCQ-MSX0932), the Natural Science Foundation of Hunan Province for Youths (Grant No. 2024JJ6210), and the Research and Innovation Team Cultivation Program of Chongqing University of Technology (Grant No. 2023TDZ007). 
\end{acknowledgments}


%\bibliography{reference.bib}

\begin{thebibliography}{99}
	
	\bibitem{EHT2019a}
	K. Akiyama \emph{et al.} (The Event Horizon Telescope Collaboration), First M87 Event Horizon Telescope Results. I. The Shadow of the Supermassive Black Hole, 
	\href{https://doi.org/10.3847/2041-8213/ab0ec7}{Astrophys. J. {\bf 875}, L1 (2019).}  \href{https://doi.org/10.48550/arXiv.1906.11238}{arXiv:1906.11238 [astro-ph.GA]}
	
	\bibitem{EHT2019b}
	K. Akiyama \textit{et al.} (The Event Horizon Telescope Collaboration),
	First M87 Event Horizon Telescope Results. IV. Imaging the Central Supermassive Black Hole,
	\href{https://doi.org/10.3847/2041-8213/ab0e85}{Astrophys. J. Lett. \textbf{875}, L4 (2019).}
	\href{https://doi.org/10.48550/arXiv.1906.11241}{arXiv:1906.11241[astro-ph.GA]}
	
	\bibitem{ETH2022}
	K. Akiyama \emph{et al.} (The Event Horizon Telescope Collaboration), First Sagittarius A* Event Horizon Telescope Results. I. The Shadow of the Supermassive Black Hole in the Center of the Milky Way, \href{https://doi.org/10.3847/2041-8213/ac6674}{Astrophys. J. Lett. {\bf 930} 2, L12 (2022).}
	
	\bibitem{LIGO2016}
	B. P. Abbott \emph{et al.} (LIGO Scientific Collaboration and Virgo Collaboration), Observation of Gravitational Waves from a Binary Black Hole Merger, 
	\href{https://doi.org/10.1103/PhysRevLett.116.061102}{Phys. Rev. Lett. {\bf 116}, 061102 (2016).}
	
	\bibitem{LIGO2016b}
	B. P. Abbott \emph{et al.} (LIGO Scientific Collaboration and Virgo Collaboration), Binary Black Hole Mergers in the first Advanced LIGO Observing Run,
	\href{https://doi.org/10.1103/PhysRevX.6.041015}{Phys. Rev. X {\bf 6}, 041015 (2016);} \href{https://doi.org/10.1103/PhysRevX.8.039903}{Erratum: Phys. Rev. X {\bf 8}, 039903 (2018).}
	
	\bibitem{Virbhadra2000}
	K. S. Virbhadra and G. F. R. Ellis,
	Schwarzschild black hole lensing,
	\href{https://doi.org/10.1103/PhysRevD.62.084003}{Phys. Rev. D \textbf{62}, 084003 (2000).}
	\href{https://doi.org/10.48550/arXiv.astro-ph/9904193}{arXiv:astro-ph/9904193}
	
	\bibitem{Bozza2002}
	V. Bozza,
	Gravitational lensing in the strong field limit,
	\href{https://doi.org/10.1103/PhysRevD.66.103001}{Phys. Rev. D \textbf{66}, 103001 (2002).}
	\href{https://doi.org/10.48550/arXiv.gr-qc/0208075}{arXiv:gr-qc/0208075}
	
	\bibitem{Iyer2006}
	S. V. Iyer and A. O. Petters,
	Light's bending angle due to black holes: From the photon sphere to infinity,
	\href{https://doi.org/10.1007/s10714-007-0481-8}{Gen. Rel. Grav. \textbf{39}, 1563-1582 (2007).}
	\href{https://doi.org/10.48550/arXiv.gr-qc/0611086}{arXiv:gr-qc/0611086}
	
	\bibitem{Tsukamoto2016}
	N. Tsukamoto,
	Deflection angle in the strong deflection limit in a general asymptotically flat, static, spherically symmetric spacetime,
	\href{https://doi.org/10.1103/PhysRevD.95.064035}{Phys. Rev. D \textbf{95}, 064035 (2017).}
	\href{https://doi.org/10.48550/arXiv.1612.08251}{arXiv:1612.08251[gr-qc]}
	
	\bibitem{Grover2017}
	J. Grover and A. Wittig,
	Black Hole Shadows and Invariant Phase Space Structures,
	\href{https://doi.org/10.1103/PhysRevD.96.024045}{Phys. Rev. D \textbf{96}, 024045 (2017).}
	\href{https://doi.org/10.48550/arXiv.1705.07061}{arXiv:1705.07061[gr-qc]}
	
	\bibitem{Tsukamoto2018}
	N. Tsukamoto, Black hole shadow in an asymptotically-flat, stationary, and axisymmetric spacetime: The Kerr-Newman and rotating regular black holes,
	\href{https://doi.org/10.1103/PhysRevD.97.064021}{Phys. Rev. D \textbf{97}, 064021 (2018).}
	\href{https://doi.org/10.48550/arXiv.1708.07427}{arXiv:1708.07427[gr-qc]}
	
	\bibitem{Perlick2022}
	V. Perlick and O. Y. Tsupko, \emph{Calculating black hole shadows: review of analytical studies}, 
	\href{https://doi.org/10.1016/j.physrep.2021.10.004}{Phys. Rep. {\bf 947}, 1-39 (2022).}  
	\href{https://doi.org/10.48550/arXiv.2105.07101}{arXiv:2105.07101[gr-qc]}
	
	\bibitem{Vagnozzi2023}
	S. Vagnozzi, R. Roy, Y. D. Tsai, L. Visinelli, M. Afrin, A. Allahyari, P. Bambhaniya, D. Dey, S. G. Ghosh and P. S. Joshi, \emph{et al.}
	Horizon-scale tests of gravity theories and fundamental physics from the Event Horizon Telescope image of Sagittarius A,
	\href{https://doi.org/10.1088/1361-6382/acd97b}{Class. Quant. Grav. \textbf{40}, 165007 (2023).}
	\href{https://doi.org/10.48550/arXiv.2205.07787}{arXiv:2205.07787[gr-qc]}
	
	\bibitem{Chen2023}
	S. Chen, J. Jing, W. -L. Qian and B. Wang, Black hole images: A Review, \href{https://doi.org/10.1007/s11433-022-2059-5}{Sci. China Phys. Mech. Astron. {\bf 66}, 260401 (2023).}
	\href{https://doi.org/10.48550/arXiv.2301.00113}{arXiv:2301.00113[astro-ph.HE]}
	
	\bibitem{Gralla2019}
	S. E. Gralla, D. E. Holz and R. M. Wald,
	Black Hole Shadows, Photon Rings, and Lensing Rings,
	\href{https://doi.org/10.1103/PhysRevD.100.024018}{Phys. Rev. D \textbf{100}, 024018 (2019).}
	\href{https://doi.org/10.48550/arXiv.1906.00873}{arXiv:1906.00873[astro-ph.HE]}
	
	\bibitem{Jaroszynski1997}
	M. Jaroszynski and A. Kurpiewski,
	Optics near kerr black holes: spectra of advection dominated accretion flows,
	\href{https://ui.adsabs.harvard.edu/abs/1997A&A...326..419J}{Astron. Astrophys. \textbf{326}, 419-426 (1997).}
	\href{https://doi.org/10.48550/10.48550/arXiv.astro-ph/9705044}{arXiv:astro-ph/9705044}
	
	\bibitem{ZengXX2020}
	X. -X. Zeng, H. -Q. Zhang, H. -B. Zhang, Shadows and photon spheres with spherical accretions in the four-dimensional Gauss–Bonnet black hole, \href{https://doi.org/10.1140/epjc/s10052-020-08449-y}{Eur. Phys. J. C \textbf{80}, 872 (2020).}
	\href{https://doi.org/10.48550/arXiv.2004.12074}{arXiv:2004.12074[gr-qc]}
	
	\bibitem{Cardoso2008}
	V. Cardoso, A. S. Miranda, E. Berti, H. Witek and V. T. Zanchin,
	Geodesic stability, Lyapunov exponents and quasinormal modes,
	\href{https://doi.org/10.1103/PhysRevD.79.064016}{Phys. Rev. D \textbf{79}, 064016 (2009).}
	\href{https://doi.org/10.48550/arXiv.0812.1806}{arXiv:0812.1806[hep-th]}
	
	\bibitem{Cardoso2014}
	V. Cardoso, L. C. B. Crispino, C. F. B. Macedo, H. Okawa and P. Pani,
	Light rings as observational evidence for event horizons: long-lived modes, ergoregions and nonlinear instabilities of ultracompact objects,
	\href{https://doi.org/10.1103/PhysRevD.90.044069}{Phys. Rev. D \textbf{90}, 044069 (2014).}
	\href{https://doi.org/10.48550/arXiv.1406.5510}{arXiv:1406.5510[gr-qc]}
	
	\bibitem{Keir2016}
	J. Keir,
	Slowly decaying waves on spherically symmetric spacetimes and ultracompact neutron stars,
	\href{https://doi.org/10.1088/0264-9381/33/13/135009}{Class. Quant. Grav. \textbf{33}, 135009 (2016).}
	\href{https://doi.org/10.48550/arXiv.1404.7036}{arXiv:1404.7036[gr-qc]}
	
	\bibitem{GaoSJ2022}
	M. Guo, Z. Zhong, J. Wang and S. Gao, Light rings and long-lived modes in quasiblack hole spacetimes,
	\href{https://doi.org/PhysRevD.105.024049}{Phys. Rev. D \textbf{105}, 024049 (2022).}
	\href{https://doi.org/10.48550/arXiv.2108.08967}{arXiv:2108.08967[gr-qc]}
	
	\bibitem{Cunha2023}
	P. V. P. Cunha, C. A. R. Herdeiro, E. Radu and N. Sanchis-Gual,
	Exotic Compact Objects and the Fate of the Light-Ring Instability,
	\href{https://doi.org/10.1103/PhysRevLett.130.061401}{Phys. Rev. Lett. \textbf{130}, 061401 (2023).}
	\href{https://doi.org/10.48550/arXiv.2207.13713}{arXiv:2207.13713[gr-qc]}
	
	\bibitem{Deich2023}
	A. Deich, N. Yunes and C. Gammie, Lyapunov exponents to test general relativity,
	\href{https://doi.org/10.1103/PhysRevD.110.044033}{Phys. Rev. D \textbf{110}, 044033 (2024).}
	\href{https://doi.org/arXiv.2308.07232}{arXiv:2308.07232[gr-qc]}
	
	\bibitem{Johannsen2013}
	T. Johannsen, Photon Rings around Kerr and Kerr-like Black Holes, 
	\href{https://doi.org/10.1088/0004-637X/777/2/170}{Astrophys. J. {\bf 777}, 170 (2013).}    
	\href{https://arxiv.org/abs/1501.02814}{arXiv:1501.02814[astro-ph.HE]}
	
	\bibitem{Hod2011}
	S. Hod, Hairy Black Holes and Null Circular Geodesics, \href{https://doi.org/10.1103/PhysRevD.84.124030}{Phys. Rev. D \textbf{84}, 124030 (2011)}   
	\href{https://doi.org/10.48550/arXiv.1112.3286}{arXiv:1112.3286[gr-qc]}
	
	\bibitem{Hod2013}
	S. Hod, Upper bound on the radii of black-hole photonspheres,
	\href{https://doi.org/10.1016/j.physletb.2013.10.047}{Phys. Lett. B \textbf{727}, 345-348 (2013).}
	\href{https://doi.org/10.48550/arXiv.1701.06587}{arXiv:1701.06587[gr-qc]}
	
	\bibitem{Mishra2019}
	A. K. Mishra, S. Chakraborty and S. Sarkar, Understanding photon sphere and black hole shadow in dynamically evolving spacetimes, \href{https://doi.org/10.1103/PhysRevD.99.104080}{Phys. Rev. D \textbf{99}, 104080 (2019).}
	\href{https://doi.org/10.48550/arXiv.1903.06376}{arXiv:1903.06376[gr-qc]}
	
	\bibitem{Wielgus2021}
	M. Wielgus, Photon rings of spherically symmetric black holes and robust tests of non-Kerr metrics, 
	\href{https://doi.org/10.1103/PhysRevD.104.124058}{Phys. Rev. D \textbf{104}, 124058 (2021).}
	\href{https://doi.org/10.48550/arXiv.2109.10840}{arXiv:2109.10840[gr-qc]}
	
	\bibitem{Raffaelli2021}	
	B. Raffaelli, Hidden conformal symmetry on the black hole photon sphere,   
	\href{https://doi.org/10.1007/JHEP03(2022)125}{J. High Energ. Phys. {\bf 2022}, 125 (2022).}
	\href{https://doi.org/10.48550/arXiv.2112.12543}{arXiv:2112.12543[gr-qc]}
	
	\bibitem{Bargueno2023}
	P. Bargue\~no, Light rings in static and extremal black holes,
	\href{https://doi.org/10.1103/PhysRevD.107.104029}{Phys. Rev. D \textbf{107}, 104029 (2023).}
	\href{https://doi.org/10.48550/arXiv.2211.16899}{arXiv:2211.16899[gr-qc]}
	
	\bibitem{Vertogradov2024}
	V. Vertogradov and A. \"Ovg\"un,
	Analyzing the influence of geometrical deformation on photon sphere and shadow radius: A new analytical approach \textemdash{} Spherically symmetric spacetimes,
	\href{https://doi.org/10.1016/j.dark.2024.101541}{Phys. Dark Univ. \textbf{45}, 101541 (2024).}
	\href{https://doi.org/10.48550/arXiv.2404.04046}{arXiv:2404.04046[gr-qc]}
	
	\bibitem{Shaikh2018}
	R. Shaikh, P. Kocherlakota, R. Narayan and P. S. Joshi, Shadows of spherically symmetric black holes and naked singularities, \href{https://doi.org/10.1093/mnras/sty2624}{Mon. Not. Roy. Astron. Soc. \textbf{482}, 52-64 (2019).}
	\href{https://doi.org/10.48550/arXiv.1802.08060}{arXiv:1802.08060[astro-ph.HE]}
	
	\bibitem{Joshi2020}
	A. B. Joshi, D. Dey, P. S. Joshi and P. Bambhaniya, Shadow of a naked singularity without photon sphere, \href{https://doi.org/10.1103/PhysRevD.102.024022}{Phys. Rev. D \textbf{102}, 024022 (2020).}   \href{https://doi.org/10.48550/arXiv.2004.06525}{arXiv:2004.06525[gr-qc]}
	
	\bibitem{Berry2020}
	T. Berry, A. Simpson and M. Visser, Photon spheres, ISCOs, and OSCOs: Astrophysical observables for regular black holes with asymptotically Minkowski cores, \href{https://doi.org/10.3390/universe7010002}{Universe \textbf{7}, 2 (2020).}
	\href{https://doi.org/10.48550/arXiv.2008.13308}{arXiv:2008.13308[gr-qc]}
	
	\bibitem{Isomura2023}
	K. Isomura, R. Suzuki and S. Tomizawa, Particle motions around regular black holes,
	\href{https://doi.org/10.1103/PhysRevD.107.084003}{Phys. Rev. D \textbf{107}, 084003 (2023).}
	\href{https://doi.org/10.48550/arXiv.2301.10465}{arXiv:2301.10465[gr-qc]}
	
	\bibitem{Murk2024}
	S. Murk and I. Soranidis, Light rings and causality for nonsingular ultracompact objects sourced by nonlinear electrodynamics,
	\href{https://doi.org/10.1103/PhysRevD.110.044064}{Phys. Rev. D {\textbf 110}, 044064 (2024).} 
	\href{https://doi.org/10.48550/arXiv.2406.07957}{arXiv:2406.07957[gr-qc].}
	
	\bibitem{Tsukamoto2024}
	N. Tsukamoto,
	Circular light orbits of a general, static, and spherical symmetrical wormhole with $Z_2$ symmetry,
	\href{https://doi.org/10.1140/epjc/s10052-024-13696-4}{Eur. Phys. J. C {\textbf 84}, 1325 (2024).}
	\href{https://doi.org/10.48550/arXiv.2401.07846}{arXiv:2401.07846[gr-qc].}
	
	\bibitem{Cunha2017}
	P. V. P. Cunha, E. Berti and C. A. R. Herdeiro, Light-Ring Stability for Ultracompact Objects,
	\href{https://doi.org/10.1103/PhysRevLett.119.251102}{Phys. Rev. Lett. {\bf 119}, 251102 (2017).}
	\href{https://doi.org/10.48550/arXiv.1708.04211}{arXiv:1708.04211[gr-qc]}
	
	\bibitem{Cunha2020}
	P. V. P. Cunha and C. A. R. Herdeiro, Stationary Black Holes and Light Rings,
	\href{https://doi.org/10.1103/PhysRevLett.124.181101}{Phys. Rev. Lett. {\bf 124}, 181101 (2020).}
	\href{https://doi.org/10.48550/arXiv.2003.06445}{arXiv:2003.06445[gr-qc]}
	
	\bibitem{WeiSW2020}
	S. W. Wei, Topological charge and black hole photon spheres, \href{https://doi.org/10.48550/arXiv.2006.02112}{Phys. Rev. D \textbf{102}, 064039 (2020).}   
	\href{https://doi.org/10.48550/arXiv.2006.02112}{arXiv:2006.02112[gr-qc]}
	
	\bibitem{LiuHS2019}
	H. S. Liu, Z. F. Mai, Y. Z. Li and H. L\"u, Quasi-topological Electromagnetism: Dark Energy, Dyonic Black Holes, Stable Photon Spheres and Hidden Electromagnetic Duality,
	\href{https://doi.org/10.1007/s11433-019-1446-1}{Sci. China Phys. Mech. Astron. \textbf{63}, 240411 (2020).}
	\href{https://doi.org/10.48550/arXiv.1907.10876}{arXiv:1907.10876[hep-th]}
	
	\bibitem{GanQY2021}	
	Q. Gan, P. Wang, H. Wu and H. Yang, Photon spheres and spherical accretion image of a hairy black hole, 
	\href{https://doi.org/10.1103/PhysRevD.104.024003}{Phys. Rev. D {\bf 104}, 024003 (2021).}   
	\href{https://arxiv.org/abs/2104.08703}{arXiv:2104.08703[gr-qc]}
	
	\bibitem{GanQY2021b}
	Q. Gan, P. Wang, H. Wu and H. Yang, Photon ring and observational appearance of a hairy black hole,
	\href{https://doi.org/10.1103/PhysRevD.104.044049}{Phys. Rev. D \textbf{104}, 044049 (2021).}
	\href{https://doi.org/10.48550/arXiv.2105.11770}{arXiv:2105.11770[gr-qc]}
	
	\bibitem{GuoGZ2023}
	G. Guo, Y. Lu, P. Wang, H. Wu and H. Yang, Black holes with multiple photon spheres,
	\href{https://doi.org/10.1103/PhysRevD.107.124037}{Phys. Rev. D \textbf{107}, 124037 (2023).}
	\href{https://doi.org/10.48550/arXiv.2212.12901}{arXiv:2212.12901[gr-qc]}
	
	\bibitem{Gibbons2016}
	M. Cvetic, G. W. Gibbons and C. N. Pope, Photon spheres and sonic horizons in black holes from supergravity and other theories, \href{https://doi.org/10.1103/PhysRevD.94.106005}{Phys. Rev. D \textbf{94}, 106005 (2016).}   
	\href{https://doi.org/10.48550/arXiv.1608.02202}{arXiv:1608.02202[gr-qc]}
	
	\bibitem{Cederbaum2015}
	C. Cederbaum and G. J. Galloway,
	Uniqueness of photon spheres via positive mass rigidity,
	\href{https://doi.org/10.4310/CAG.2017.v25.n2.a2}{Commun. Anal. Geom. \textbf{25}, 303-320 (2017).}
	\href{https://doi.org/10.48550/arXiv.1504.05804}{arXiv:1504.05804[math.DG]}
	
	\bibitem{Cederbaum2016}
	C. Cederbaum and G. J. Galloway, Uniqueness of photon spheres in electro-vacuum spacetimes,
	\href{https://doi.org/10.1088/0264-9381/33/7/075006}{Class. Quant. Grav. \textbf{33}, 075006 (2016).}
	\href{https://doi.org/10.48550/arXiv.1508.00355}{arXiv:1508.00355[math.DG]}
	
	\bibitem{JiaJJ2018a}
	J. Jia, J. Liu, X. Liu, Z. Mo, X. Pang, Y. Wang and N. Yang, Existence and stability of circular orbits in general static and spherically symmetric spacetimes, \href{https://doi.org/10.1007/s10714-017-2337-1}{Gen. Rel. Grav. {\bf 50}, 17 (2018).}
	\href{https://doi.org/10.48550/arXiv.1702.05889}{arXiv:1702.05889[gr-qc]}
	
	\bibitem{JiaJJ2018b}
	J. Jia, X. Pang and N. Yang, Existence and stability of circular orbits in static and axisymmetric spacetimes, \href{https://doi.org/10.1007/s10714-018-2364-6}{Gen Relativ Gravit {\bf 50}, 41 (2018).} 
	\href{https://doi.org/10.48550/arXiv.1704.01689}{arXiv:1704.01689[gr-qc]}
	
	\bibitem{GaoSJ2021}
	M. Guo and S. Gao, Universal Properties of Light Rings for Stationary Axisymmetric Spacetimes,
	\href{https://doi.org/10.1103/PhysRevD.103.104031}{Phys. Rev. D \textbf{103}, 104031 (2021).}
	\href{https://doi.org/10.48550/arXiv.2011.02211}{arXiv:2011.02211[gr-qc]}
	
	\bibitem{Ghosh2021}
	R. Ghosh and S. Sarkar, Light rings of stationary spacetimes,
	\href{https://doi.org/10.1103/PhysRevD.104.044019}{Phys. Rev. D {\bf 104}, 044019 (2021).}
	\href{https://doi.org/10.48550/arXiv.2107.07370}{arXiv:2107.07370[gr-qc]} 	
	
	\bibitem{Virbhadra2001}
	Clarissa-Marie Claudel, K. S. Virbhadra and G. F. R. Ellis, The geometry of photon surfaces,
	\href{https://doi.org/10.1063/1.1308507}{J. Math. Phys. {\bf 42}, 818-838 (2001).}
	\href{https://doi.org/10.48550/arXiv.gr-qc/0005050}{arXiv:0005050[gr-qc]}
	
	\bibitem{Koga2019}
	Y. Koga and T. Harada, Stability of null orbits on photon spheres and photon surfaces, \href{https://doi.org/10.1103/PhysRevD.100.064040}{Phys. Rev. D \textbf{100}, 064040 (2019).}   \href{https://doi.org/10.48550/arXiv.1907.07336}{arXiv:1907.07336[gr-qc]}
	
	\bibitem{Kobialko2022}
	K. Kobialko, I. Bogush and D. Gal'tsov, Geometry of massive particle surfaces,
	\href{https://doi.org/10.1103/PhysRevD.106.084032}{Phys. Rev. D \textbf{106}, 084032 (2022).}
	\href{https://doi.org/10.48550/arXiv.2208.02690}{arXiv:2208.02690[gr-qc]}
	
	\bibitem{SongY2023}
	Y. Song, Y. Cen, L. Tang, J. Hu, K. Diao, X. Zhao and S. Shi,
	The particle surface of spinning test particles,
	\href{https://doi.org/10.1140/epjc/s10052-023-11970-5}{Eur. Phys. J. C \textbf{83}, 833 (2023).}
	\href{https://doi.org/10.48550/arXiv.2208.03665}{arXiv:2208.03665[gr-qc]}
	
	\bibitem{Cunha2018}
	P. V. P. Cunha and C. A. R. Herdeiro, Shadows and strong gravitational lensing: a brief review, \href{https://doi.org/10.1007/s10714-018-2361-9}{Gen Relativ Gravit {\bf 50}, 42 (2018).}
	\href{https://doi.org/10.48550/arXiv.1801.00860}{arXiv:1801.00860[gr-qc]}
	
	\bibitem{DuanYS}
	Y. S. Duan and M. L. Ge,
	SU(2) Gauge Theory and Electrodynamics with N Magnetic Monopoles,
	\href{https://doi.org/10.1142/9789813237278_0001}{Sci. Sin. \textbf{9}, 1072 (1979).}
	
	\bibitem{WeiSW2023}
	S. W. Wei and Y. X. Liu, Topology of equatorial timelike circular orbits around stationary black holes, \href{https://doi.org/10.1103/PhysRevD.107.064006}{Phys. Rev. D {\bf 107}, 064006 (2023).}
	\href{https://doi.org/10.48550/arXiv.2207.08397}{arXiv:2207.08397[gr-qc]}
	
	\bibitem{WeiSW2023b}
	X. Ye and S. W. Wei, Distinct topological configurations of equatorial timelike circular orbit for spherically symmetric (hairy) black holes, \href{https://doi.org/10.1088/1475-7516/2023/07/049}{J. Cosmol. Astropart. Phys. \textbf{2023(07)}, 049 (2023).}
	\href{https://doi.org/10.48550/arXiv.2301.04786}{arXiv:2301.04786[gr-qc]}
	
	\bibitem{WeiSW2023c}
	S. W. Wei, Y. P. Zhang, Y. X. Liu and R. B. Mann, Static spheres around spherically symmetric black hole spacetime,
	\href{https://doi.org/10.1103/PhysRevResearch.5.043050}{Phys. Rev. Res. \textbf{5}, 043050 (2023).}
	\href{https://doi.org/10.48550/arXiv.2303.06814}{arXiv:2303.06814[gr-qc]}
	
	\bibitem{Lima2022}
	H. C. D. Lima Junior, J. Z. Yang, L. C. B. Crispino, P. V. P. Cunha and C. A. R. Herdeiro,
	Einstein-Maxwell-dilaton neutral black holes in strong magnetic fields: Topological charge, shadows, and lensing,
	\href{https://doi.org/10.1103/PhysRevD.105.064070}{Phys. Rev. D \textbf{105}, 064070 (2022).}
	\href{https://doi.org/10.48550/arXiv.2112.10802}{arXiv:2112.10802[gr-qc]}
	
	\bibitem{YinJ2023}
	J. Yin, J. Jiang and M. Zhang, Kinematic topologies of black holes,
	\href{https://doi.org/10.1103/PhysRevD.108.044077}{Phys. Rev. D \textbf{108}, 044077 (2023).}
	\href{https://doi.org/10.48550/arXiv.2305.14179}{arXiv:2305.14179[gr-qc]}
	
	\bibitem{Tavlayan2023}
	A. Tavlayan and B. Tekin,
	Light rings around five dimensional stationary black holes and naked singularities,
	\href{https://doi.org/10.1103/PhysRevD.107.024016}{Phys. Rev. D \textbf{107}, 024016 (2023).}
	\href{https://doi.org/10.48550/arXiv.2209.14873}{arXiv:2209.14873[gr-qc]}
	
	\bibitem{Sadeghi2024}
	J. Sadeghi, M. A. S. Afshar, S. N. Gashti and M. R. Alipour, Thermodynamic topology and photon spheres in the hyperscaling violating black holes, \href{https://doi.org/10.1016/j.astropartphys.2023.102920}{Astropart.Phys. \textbf{156}  102920 (2024).}   \href{https://doi.org/10.48550/arXiv.2307.12873}{arXiv:2307.12873[gr-qc]}
	
	\bibitem{Cunha2024}
	P. V. P. Cunha, C. A. R. Herdeiro and J. P. A. Novo, Light rings on stationary axisymmetric spacetimes: Blind to the topology and able to coexist,
	\href{https://doi.org/10.1103/PhysRevD.109.064050}{Phys. Rev. D \textbf{109}, no.6, 064050 (2024).}
	\href{https://doi.org/10.48550/arXiv.2401.05495}{arXiv:2401.05495[gr-qc]}
	
	\bibitem{Xavier2024}
	S. V. M. C. B. Xavier, C. A. R. Herdeiro and L. C. B. Crispino, Traversable wormholes and light rings,
	\href{https://doi.org/10.1103/PhysRevD.109.124065}{Phys. Rev. D \textbf{109}, 124065 (2024).}
	\href{https://doi.org/10.48550/arXiv.2404.02208}{arXiv:2404.02208[gr-qc]}
	
	\bibitem{Afshar2024}
	J. Sadeghi and M. A. S. Afshar,
	The role of topological photon spheres in constraining the parameters of black holes,
	\href{https://doi.org/10.1016/j.astropartphys.2024.102994}{Astropart. Phys. \textbf{162}, 102994 (2024).}
    \href{https://doi.org/10.48550/arXiv.2405.06568}{arXiv:2405.06568[gr-qc]}
	
	\bibitem{Afshar2024b}
	J. Sadeghi and M. A. S. Afshar,
	Effective Potential and Topological Photon Spheres: A Novel Approach to Black Hole Parameter Classification,
	\href{https://doi.org/10.48550/arXiv.2405.18798}{arXiv:2405.18798[gr-qc]}
	
	\bibitem{WeiSW2022b}
	S. W. Wei and Y. -X. Liu, Topology of black hole thermodynamics, \href{https://doi.org/10.1103/PhysRevD.105.104003}{Phys. Rev. D {\bf 105}, 104003 (2022).}  
	\href{https://doi.org/10.48550/arXiv.2112.01706}{arXiv:2112.01706[gr-qc]}
	
	\bibitem{WeiSW2022}
	S. W. Wei, Y. -X. Liu and R. B. Mann, Black Hole Solutions as Topological Thermodynamic Defects, \href{https://doi.org/10.1103/PhysRevLett.129.191101}{Phys. Rev. Lett. {\bf 129}, 191101 (2022).}   \href{https://doi.org/10.48550/arXiv.2208.01932}{arXiv:2208.01932[gr-qc]}
	
	\bibitem{WeiSW2024}
	S. W. Wei, Y. X. Liu and R. B. Mann,
	Universal topological classifications of black hole thermodynamics,
	\href{https://doi.org/10.1103/PhysRevD.110.L081501}{Phys. Rev. D \textbf{110}, L081501 (2024).}
	\href{https://doi.org/10.48550/arXiv.2409.09333}{arXiv:2409.09333[gr-qc]}
	
	\bibitem{WuD2024}
	X. D. Zhu, W. Liu and D. Wu,
	Universal thermodynamic topological classes of rotating black holes,
	\href{https://doi.org/10.48550/arXiv.2409.12747}{arXiv:2409.12747[hep-th]}
	
    \bibitem{QiaoCK2022a}
    C. K. Qiao and M. Li, A Geometric Approach on Circular Photon Orbits and Black Hole Shadow, \href{https://doi.org/10.1103/PhysRevD.106.L021501}{Phys. Rev. D {\bf 106}, L021501 (2022).} \href{https://doi.org/10.48550/arXiv.2204.07297}{arXiv:2204.07297[qr-qc]}
    
    \bibitem{QiaoCK2022b}
    C. K. Qiao, Curvatures, photon spheres, and black hole shadows, \href{https://doi.org/10.1103/PhysRevD.106.084060}{Phys. Rev. D {\bf 106}, 084060 (2022).}   \href{https://doi.org/10.48550/arXiv.2208.01771}{arXiv:2208.01771[gr-qc]}
    
    \bibitem{Cunha2022}
    P. V. P. Cunha, C. A. R. Herdeiro and J. P. A. Novo, Null and timelike circular orbits from equivalent 2D metrics, 
    \href{https://doi.org/10.1088/1361-6382/ac987e}{Classical Quantum Gravity {\bf 39}, 225007 (2022).} \href{https://doi.org/10.48550/arXiv.2207.14506}{arXiv:2207.14506[gr-qc]}
    
    \bibitem{Bermudez-Cardenas2024}
    B. Berm\'udez-C\'ardenas and O. L. Andino,
    On massive particle surfaces, partial umbilicity and circular orbits,
    \href{https://doi.org/10.48550/arXiv.2409.10789}{arXiv:2409.10789[gr-qc]}
    
    \bibitem{Gallo2024}
    E. Gallo and T. M\"adler,
    Bounds for Lyapunov exponent of circular light orbits in black holes,
    \href{https://doi.org/10.1140/epjc/s10052-025-14046-8}{Eur. Phys. J. C \textbf{85}, 299 (2025).}
    \href{https://doi.org/10.48550/arXiv.2412.10328}{arXiv:2412.10328[gr-qc]}
    
    \bibitem{QiaoCK2024}
    C. K. Qiao, The existence and distribution of photon spheres near spherically symmetric black holes: a geometric analysis,
    \href{https://doi.org/10.1140/epjc/s10052-025-13912-9}{Eur. Phys. J. C {\bf 85}, 191 (2025).}
    \href{https://doi.org/10.48550/arXiv.2407.14035}{arXiv:2407.14035[gr-qc].}
	
	\bibitem{Abramowicz1988}
	M. A. Abramowicz, B. Carter and J. P. Lasota, Optical reference geometry for stationary and static dynamics, 
	\href{https://doi.org/10.1007/BF00758937}{Gen. Relativ. Gravit. {\bf 20}, 1173–1183 (1988).}
	
	\bibitem{Gibbons2009}
	G. W. Gibbons and C. M. Warnick, Universal properties of the near-horizon optical geometry, 
	\href{https://doi.org/10.1103/PhysRevD.79.064031}{Phys. Rev. D {\bf 79}, 064031 (2009).}  
	\href{https://doi.org/10.48550/arXiv.0809.1571}{arXiv:0809.1571[gr-qc]}
	
	\bibitem{Gibbons2008}
	G. W. Gibbons and M. C. Werner, Applications of the Gauss-Bonnet theorem to gravitational lensing, 
	\href{https://doi.org/10.1088/0264-9381/25/23/235009}{Classical Quantum Gravity {\bf 25}, 235009 (2008).}  \href{https://doi.org/10.48550/arXiv.0807.0854}{arXiv:0807.0854[gr-qc]}
	
	\bibitem{Ishihara2016a}	
	A. Ishihara, Y. Suzuki, T. Ono, T. Kitamura, and H. Asada, Gravitational bending angle of light for finite distance and the Gauss-Bonnet theorem, \href{https://doi.org/10.1103/PhysRevD.94.084015}{Phys. Rev. D {\bf 94}, 084015 (2016).} 
	\href{https://doi.org/10.48550/arXiv.1604.08308}{arXiv:1604.08308[gr-qc]}
	
	\bibitem{Ishihara2016b}	
	A. Ishihara, Y. Suzuki, T. Ono and H. Asada, Finite-distance corrections to the gravitational bending angle of light in the strong deflection limit, \href{https://doi.org/10.1103/PhysRevD.95.044017}{Phys. Rev. D {\bf 95}, 044017 (2017).}  
	\href{https://doi.org/10.48550/arXiv.1612.04044}{arXiv:1612.04044[gr-qc]}
	
	\bibitem{Werner2012}	
	M. C. Werner, Gravitational lensing in the Kerr-Randers optical geometry, 
	\href{https://doi.org/10.1007/s10714-012-1458-9}{Gen. Relativ. Gravit. {\bf 44}, 3047-3057 (2012).}  
	\href{https://doi.org/10.48550/arXiv.1205.3876}{arXiv:1205.3876[gr-qc]}
	
	\bibitem{Asida2017}
	T. Ono, A. Ishihara, H. Asada, Gravitomagnetic bending angle of light with finite-distance corrections in stationary axisymmetric spacetimes,
	\href{https://doi.org/10.1103/PhysRevD.96.104037}{Phys. Rev. D \textbf{96}, 104037 (2017).}
	\href{https://doi.org/10.48550/arXiv.1704.05615}{arXiv:1704.05615[gr-qc]} 
	
	\bibitem{Jusufi2017}
	K. Jusufi and A. \"Ovg\"un,
	Gravitational Lensing by Rotating Wormholes,
	\href{https://doi.org/10.1103/PhysRevD.97.024042}{Phys. Rev. D \textbf{97}, 024042 (2018).}
	\href{https://doi.org/10.48550/arXiv.1708.06725}{arXiv:1708.06725[gr-qc]}
	
	\bibitem{LiZH2019a}
	Z. Li and J. Jia,
	The finite-distance gravitational deflection of massive particles in stationary spacetime: a Jacobi metric approach,''
	\href{https://doi.org/10.1140/epjc/s10052-020-7665-8}{Eur. Phys. J. C \textbf{80}, 157 (2020).}
	\href{https://doi.org/10.48550/arXiv.1912.05194}{arXiv:1912.05194 [gr-qc]}
	
	\bibitem{LiZH2019b}
	Z. Li and T. Zhou,
	Equivalence of Gibbons-Werner method to geodesics method in the study of gravitational lensing,
	\href{https://doi.org/10.1103/PhysRevD.101.044043}{Phys. Rev. D \textbf{101}, 044043 (2020).}
	doi:10.1103/PhysRevD.101.044043
	\href{https://doi.org/10.48550/arXiv.1908.05592}{arXiv:1908.05592[gr-qc]}
	
	\bibitem{Crisnejo2018}
	G. Crisnejo and E. Gallo,
	Weak lensing in a plasma medium and gravitational deflection of massive particles using the Gauss-Bonnet theorem. A unified treatment,
	\href{https://doi.org/10.1103/PhysRevD.97.124016}{Phys. Rev. D \textbf{97}, 124016 (2018).}
	\href{https://doi.org/10.48550/arXiv.1804.05473}{arXiv:1804.05473[gr-qc]}
	
	\bibitem{Ono2019}
	T. Ono and H. Asada, The effects of finite distance on the gravitational deflection angle of light,
	\href{https://doi.org/10.3390/universe5110218}{Universe {\bf 5}, 218 (2019).}
	\href{https://doi.org/10.48550/arXiv.1906.02414}{arXiv:1906.02414[gr-qc]}
	
	\bibitem{Takizawa2020}	
	K. Takizawa, T. Ono, and H. Asada, Gravitational deflection angle of light: Definition by an observer and its application to an asymptotically nonflat spacetime, \href{https://doi.org/10.1103/PhysRevD.101.104032}{Phys. Rev. D {\bf 101}, 104032 (2020).}  \href{https://arxiv.org/abs/2001.03290}{arXiv:2001.03290[gr-qc]}
	
	\bibitem{LiZH2020}	
	Z. Li, G. Zhang and A. \"Ovg\"un, Circular orbit of a particle and weak gravitational lensing, 
	\href{https://doi.org/10.1103/PhysRevD.101.124058}{Phys. Rev. D {\bf 101}, 124058 (2020).}  \href{https://arxiv.org/abs/2006.13047}{arXiv:2006.13047[gr-qc]}
	
	\bibitem{HuangY2022}
	Y. Huang and Z. Cao, Generalized Gibbons-Werner method for deflection angle, \href{https://doi.org/10.1103/PhysRevD.106.104043}{Phys. Rev. D {\bf 106}, 104043 (2022).}
	
	\bibitem{HuangY2023}
	Y. Huang, Z. Cao and Z. Lu, Generalized Gibbons-Werner method for stationary spacetimes, \href{https://doi.org/10.1088/1475-7516/2024/01/013}{JCAP {\bf 2024(01)}, 013 (2024).}   \href{https://doi.org/10.48550/arXiv.2306.04145}{arXiv:2306.04145[gr-qc]}
	
	\bibitem{Takizawa2023}
	K. Takizawa and H. Asada, Gravitational lens on a static optical constant-curvature background: Its application to the Weyl gravity model,
	\href{https://doi.org/10.1103/PhysRevD.108.104055}{Phys. Rev. D {\bf 108}, 104055 (2023).}
	\href{https://doi.org/10.48550/arXiv.2304.02219}{arXiv:2304.02219[gr-qc]}
	
	\bibitem{ZhangZ2021}
	Z. Zhang,
	Geometrization of light bending and its application to SdS$_{w}$ spacetime,
	\href{https://doi.org/10.1088/1361-6382/ac38d1}{Class. Quant. Grav. \textbf{39}, 015003 (2022).}
	\href{https://doi.org/10.48550/arXiv.2112.04149}{arXiv:2112.04149[gr-qc]}
	
	\bibitem{ZhangZ2024}
	Z. Zhang and R. Zhang,
	On the global Gaussian bending measure and its applications in stationary spacetimes,
	\href{https://doi.org/10.48550/arXiv.2408.02195}{arXiv:2408.02195[gr-qc].}
    
    \bibitem{Berger1988}	
    M. Berger and B. Gostiaux, \emph{Differential Geometry: Manifolds, Curves, and Surfaces}, Springer-Verlag, New York (1988). %\href{https://doi.org/10.1007/978-1-4612-1033-7}{ISBN: 978-0-387-96626-7}
    
    % \bibitem{Berger2003}
    % M. Berger, \emph{A Panoramic View of Riemannian Geometry}, Springer-Verlag, Berlin (2003).  %ISBN:3-540-65317-1
    
    % \bibitem{Carmo1976}
    % M. Do Carmo, \emph{Differential Geometry of Curves and Surfaces}, Prentice-Hall (1976). %ISBN: 9780132125895
    
    %\bibitem{ChernWH}
    %W. -H. Chern, \emph{Differential Geometry}, Peking University Press, Beijing (2006). %ISBN: 978-7-301-10709-6 
    
    % \bibitem{Chern1999}
    % S. -S. Chern, W. -H. Chen and K. S. Lam, \emph{Lectures on Differential Geometry}, Prentice-Hall, World Scientific Publishing (1999). 
    
    \bibitem{Penrose1969}
    R. Penrose, Gravitational collapse: The role of general relativity, \href{https://doi.org/10.1023/A:1016578408204}{Revistas del Nuovo Cimento \textbf{1}, 252-276 (1969).} reprinted in \href{https://link.springer.com/article/10.1023/A:1016578408204}{Gen. Rel. Grav. \textbf{34}, 1141-1165 (2002).}
    
    \bibitem{Hawking1973}
    S. W. Hawking and G. F. R. Ellis, \emph{The Large Scale Structure of Space Time}, Cambridge University Press, Cambridge (1973)
    
    \bibitem{Wald1998}
    R. M. Wald,
    Gravitational collapse and cosmic censorship, chapter 5 in book \emph{Black Holes, Gravitational Radiation and the Universe}, edited by B. R. Iyer and B. Bhawal, pp 69-86, Springer-Verlag, Dordrecht (1998).
    % doi:10.1007/978-94-017-0934-7\_5
    \href{https://doi.org/10.48550/arXiv.gr-qc/9710068}{arXiv:gr-qc/9710068} 
    
    \bibitem{Johnson2019}
    M. D. Johnson \textit{et al.},
    Universal interferometric signatures of a black hole\textquoteright{}s photon ring,
    \href{https://doi.org/10.1126/sciadv.aaz1310}{Sci. Adv. \textbf{6}, eaaz1310 (2020).}
    \href{https://doi.org/10.48550/arXiv.1907.04329}{arXiv:1907.04329[astro-ph.IM]} 
    
    \bibitem{Gralla2020}
    S. E. Gralla, A. Lupsasca and D. P. Marrone,
    The shape of the black hole photon ring: A precise test of strong-field general relativity,
    \href{https://doi.org/10.1103/PhysRevD.102.124004}{Phys. Rev. D \textbf{102}, 124004 (2020).}
    \href{https://doi.org/10.48550/arXiv.2008.03879}{arXiv:2008.03879[gr-qc]}
    
    \bibitem{Chael2021}
    A. Chael, M. D. Johnson and A. Lupsasca,
    Observing the Inner Shadow of a Black Hole: A Direct View of the Event Horizon,''
    \href{https://doi.org/10.3847/1538-4357/ac09ee}{Astrophys. J. \textbf{918}, 6 (2021).}
    \href{https://doi.org/10.48550/arXiv.2106.00683}{arXiv:2106.00683[astro-ph.HE]}
    
    \bibitem{Broderick2022}
    A. E. Broderick \textit{et al.},
    The Photon Ring in M87*,
    \href{https://doi.org/10.3847/1538-4357/ac7c1d}{Astrophys. J. \textbf{935}, 61 (2022).}
    \href{https://doi.org/10.48550/arXiv.2208.09004}{arXiv:2208.09004[astro-ph.HE]}
    
    \bibitem{Olmo2023}
    G. J. Olmo, J. L. Rosa, D. Rubiera-Garcia and D. Saez-Chillon Gomez,
    Shadows and photon rings of regular black holes and geonic horizonless compact objects,
    \href{https://doi.org/10.1088/1361-6382/aceacd}{Class. Quant. Grav. \textbf{40}, 174002 (2023).}
    \href{https://doi.org/10.48550/arXiv.2302.12064}{arXiv:2302.12064[gr-qc]}
    
    \bibitem{Tiede2022}
    P. Tiede, M. D. Johnson, D. W. Pesce, D. C. M. Palumbo, D. O. Chang and P. Galison,
    Measuring Photon Rings with the ngEHT,
    \href{https://doi.org/10.3390/galaxies10060111}{Galaxies \textbf{10}, 111 (2022).}
    \href{https://doi.org/10.48550/arXiv.2210.13498}{arXiv:2210.13498[astro-ph.HE]}
    
    \bibitem{Lupsasca2024}
    A. Lupsasca, A. C\'ardenas-Avenda\~no, D. C. M. Palumbo, M. D. Johnson, S. E. Gralla, D. P. Marrone, P. Galison, P. Tiede and L. Keeble,
    The Black Hole Explorer: photon ring science, detection, and shape measurement,
    \href{https://doi.org/10.1117/12.3019437}{Proc. SPIE Int. Soc. Opt. Eng. \textbf{13092}, 130926Q (2024).}
    \href{https://doi.org/10.48550/arXiv.2406.09498}{arXiv:2406.09498[gr-qc]}
    
    
    
    \bibitem{ChernSS}
    D. Bao, S. S. Chern, Z. Shen, \emph{An Introduction to Riemann-Finsler Geometry}, Graduate Texts in Mathematics (GTM, volume 200), Springer, New York (2000).
    
    %\bibitem{ShenYB}
    %Y. B. Shen and Z. Shen, \emph{Introduction to Modern Finsler Geometry}, Higher Education Press, Beijing (2016). 
    
    \bibitem{Andino2019}
    M. A. Arga\~naraz and O. L. Andino,
    A Riemannian geometric approach for timelike and null spacetime geodesics,
    \href{https://doi.org/10.1007/s10714-024-03314-9}{Gen. Rel. Grav. \textbf{56}, 121 (2024).}
    \href{https://doi.org/10.48550/arXiv.2112.10910}{arXiv:2112.10910[gr-qc]}
    
    \bibitem{Andino2024}
    B.~Berm\'udez-C\'ardenas and O.~L.~Andino,
    Massive particle surfaces, partial umbilicity, and circular orbits,
    \href{https://doi.org/10.1103/PhysRevD.111.064001}{Phys. Rev. D \textbf{111}, 064001 (2025).}
    \href{https://doi.org/10.48550/arXiv.2409.10789}{arXiv:2409.10789[gr-qc]}
    
    \bibitem{Andino2025}
    B. Berm\'udez-C\'ardenas and O. L. Andino,
    Massive particle surfaces and black hole shadows from intrinsic curvature,
    \href{https://doi.org/10.48550/arXiv.2503.21203}{arXiv:2503.21203[gr-qc]}
    
    \bibitem{Shimada2010}
    J. Itoh, S. V. Sabau and H. Shimada, A Gauss-Bonnet-type formula on Riemann-Finsler surfaces with nonconstant indicatrix volume, 
    \href{https://doi.org/10.1215/0023608X-2009-008}{Kyoto J. Math. \textbf{50}, 165-192 (2010).}
    
    \bibitem{QiaoCK2025b}
    C. Qiao, M. Li, D. Xie, and M. Guo, Geometric Approach to Light Rings in Axially Symmetric Spacetimes,
    \href{https://doi.org/10.48550/arXiv.2512.20802}{arXiv:2512.20802[gr-qc]}
	
\end{thebibliography}

\end{document}